\documentclass[aps,floatfix,twocolumn,superscriptaddress,showpacs,letter]{revtex4-1}

\usepackage{float}
\usepackage{amsmath}
\usepackage{amssymb}
\usepackage{graphicx}
\usepackage{bm}
\usepackage{epic}
\usepackage{eepic}
\usepackage{pifont}
\usepackage[utf8]{inputenc}
\usepackage{rotating}
\usepackage{color}
\usepackage{nicefrac}
\usepackage{ulem}
\usepackage[caption=false]{subfig}
\usepackage{array}
\usepackage{tabularx}
\usepackage{soul}
\usepackage{hyperref}

\begin{document}
\title{Mass transfer from a sheared spherical rigid capsule}
\author{Clément Bielinski}
\affiliation{Biomechanics and Bioengineering Laboratory,\\
CNRS - Université de Technologie de Compiègne,\\
60200 Compiègne, France}
\author{Lumi Xia}
\affiliation{Biomechanics and Bioengineering Laboratory,\\
CNRS - Université de Technologie de Compiègne,\\
60200 Compiègne, France}
\author{Guillaume Helbecque}
\affiliation{Biomechanics and Bioengineering Laboratory,\\
CNRS - Université de Technologie de Compiègne,\\
60200 Compiègne, France}
\author{Badr Kaoui}
\email{badr.kaoui@utc.fr}
\affiliation{Biomechanics and Bioengineering Laboratory,\\
CNRS - Université de Technologie de Compiègne,\\
60200 Compiègne, France}
\date{\today}
\begin{abstract}
Solute mass transfer from a spherical fluid-filled rigid capsule subjected to shear flow is studied numerically, while considering unsteady, continuous and nonuniform boundary conditions on its surface.
Here, the capsule acts as a reservoir with its initially encapsulated solute concentration decaying over time.
This scenario differs from the classical case study of either constant concentration or constant mass flux at the surface of the particle.
The flow and the concentration field are computed using fully three-dimensional lattice Boltzmann simulations, where the fluid-structure two-way coupling is achieved by the immersed boundary method.
Effects of the flow and the boundary conditions on mass transfer efficacy are quantified by the Sherwood number (the dimensionless mass transfer coefficient), which is found to increase due to the combined effects of forced convection and local shear induced by the rotation of the capsule.
Having continuity of both the concentration and the mass flux on the capsule significantly decreases the Sherwood number as compared to the case with constant and uniform boundary conditions.
All the obtained results can be applied to heat transfer in case of cooling an initially hot spherical particle, for which the concentration must be replaced by the temperature and the Sherwood number by the Nusselt number.
\end{abstract}
\maketitle
\section{Problem statement}
Mass transfer from particle suspensions under flow conditions is encountered in many natural phenomena and industrial processes.
It has motivated multiple applied, as well as fundamental studies.
For example, mass transfer from a spherical particle subjected to shear flow has been studied theoretically by Acrivos and co-workers \cite{Frankel1968,Acrivos1971,Poe1976} in the limit of small Reynolds numbers (${\rm Re}\rightarrow 0$), and at either low Péclet numbers (${\rm Pe} \rightarrow 0$) or high Péclet numbers (${\rm Pe} \rightarrow \infty$).
They have derived analytically asymptotic relations for the Sherwood number ($\rm Sh$) or the Nusselt number ($\rm Nu$), which are respectively the mass and the heat transfer dimensionless coefficients.
Batchelor \cite{Batchelor1979} has obtained similar correlations using a different approach.
Later, Polyanin and Dil'man \cite{Polyanin1985} used an advanced fitting procedure to bridge the gap between the correlations of Acrivos, which are derived in the limits of ${\rm Pe} \rightarrow 0$ and ${\rm Pe} \rightarrow \infty$.
Subramanian and Koch \cite{Subramanian2006} have extended these works to finite Reynolds numbers and to report inertia effect.
Longest and Kleinstreuer \cite{Longest2004} have investigated the effect of the walls using numerical simulations.
They have proposed Sherwood number correlations as a function of the blockage ratio.
Recently, Wang and Brasseur \cite{Wang2019} studied numerically mass transfer from a freely rotating sphere in shear flow and have proposed correlations valid over a wide range of the Péclet number.
All these studies, and others summarized in Ref.~\cite{Clift1978}, are limited to particles whose surface is maintained at either constant concentration or constant mass flux.

In the present article, mass transfer from a spherical particle freely suspended in shear flow is studied numerically, while considering unsteady, continuous and nonuniform boundary conditions.
Under these conditions a particle acts as a reservoir with an initial solute load that decays over time.
Such scenario is known for oxygen and carbon dioxide transport by red blood cells, and it is increasingly used in medicine for drug delivery by particles and in food and cosmetic industry for controlled release of chemical substances using capsules.
Design of such particles with desired capacity of encapsulation and release rate present technological challenges in absence of models and studies that take into account the exact conditions under which these particles operate.
\begin{figure*}
\centering
\includegraphics[width = \textwidth]{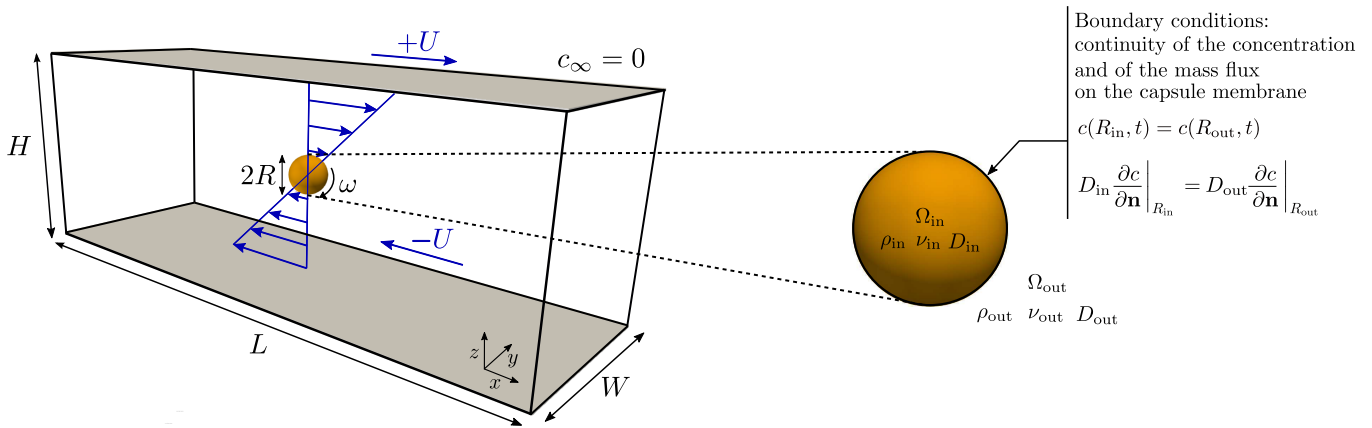}
\caption{Numerical setup of the problem. A spherical rigid capsule of radius $R$ is suspended in the center of a channel of length $L = 30R$, width $W = 10R$ and height $H = 10R$. The upper and lower walls translate in the $x$-direction at a velocity $+U$ and $-U$, respectively. This generates a simple linear shear flow that triggers the rotation of the capsule with respect to the $y$-axis at an angular velocity $\omega$. Subscripts ``${\rm in}$'' and ``${\rm out}$'' refer respectively to the properties of the inner encapsulated fluid $\Omega _{\rm in}$, and the outer surrounding fluid $\Omega _{\rm out}$ of the capsule. Continuity of both the concentration and the mass flux is considered at the membrane, in contrast to the constant boundary conditions largely used in the literature.}
\label{fig:numerical_setup}
\end{figure*}

One of the author has proposed numerical methods based on the lattice Boltzmann method to study solute release from steady and moving particles with unsteady and nonuniform boundary conditions at their surface.
He has considered both particles with zero-thickness membrane \cite{Kaoui2018,Kaoui2020} and with thick shell \cite{Kaoui2017}.
This latter has then been recently extended to cover a larger range of the Reynolds number and to highlight the contribution of the shell solute permeability on mass transfer coefficient correlation \cite{Bielinski2021}.
Here, these studies are extended to fully three-dimensional case of a neutrally-buoyant and free-torque spherical rigid capsule rotating under simple shear flow, see Fig.~\ref{fig:numerical_setup}.
The present study focuses on the effects of the shear flow and the unsteady, continuous and nonuniform boundary conditions on mass transfer efficacy.
The flow and the concentration field are computed with fully three-dimensional lattice Boltzmann simulations, where two-way fluid-structure coupling is achieved using the immersed boundary method.
The capsule membrane is considered to be of zero thickness and of infinite permeability to solute, and to be impermeable to the solvent.
The present used method and the obtained results can be applied to the case of heat transfer as well thanks to the mass/heat transfer analogy.

The article is organized as follows.
The problem setup and the mathematical formulation are introduced in Section \ref{sec:numerical_setup_maths}.
Results and discussions are reported in Section \ref{sec:results_discussions}, where the effects of the flow and the boundary conditions on mass transfer are quantified and analyzed.
Conclusions are given in Section \ref{sec:conclusion}.
The details about the numerical method and its validation are given in the Appendix.
\section{Mathematical formulation}
\label{sec:numerical_setup_maths}
\subsection{Problem setup}
Figure~\ref{fig:numerical_setup} shows the numerical setup of the problem.
A non-Brownian neutrally-buoyant spherical rigid capsule of radius $R$ is placed at the center of a channel of length $L = 30R$ and of equal width and height $W = H = 10R$.
The top and bottom walls translate along the $x$-axis at constant velocity $U$, but in opposite directions so that they generate simple linear shear flow.
Because the center of mass of the capsule is set at the zero velocity plane of the flow, the capsule will not translate.
It can only rotate with an angular velocity $\omega$.
The capsule is initially loaded with a uniform solute concentration $c_0$, while the surrounding medium is free from solute.
This establishes a concentration gradient that triggers unsteady solute diffusion from the capsule towards the bulk.
The initial load is not sustained, but it decays over time.
Continuity of both the concentration and the mass flux at the membrane of the capsule is considered in this study, and how it alters the mass transfer efficacy in comparison with the constant surface concentration is reported.
\subsection{Governing equations}
The encapsulated and the external fluids are assumed to be incompressible Newtonian fluids, of the same type, and their flow is described by the Navier-Stokes equations,
\begin{equation}
\frac{\partial {\bf u}}{\partial t} + {\bf u} \cdot \nabla {\bf u} = -\frac{\nabla p}{\rho} + {\bf f} + \nu \nabla ^2 {\bf u},
\label{eq:NS1}
\end{equation}
\begin{equation}
\nabla \cdot {\bf u} = 0,
\label{eq:NS2}
\end{equation}
where ${\bf u} = {\bf u}({\bf r},t)$ is the local fluid velocity, ${\bf f} = {\bf f}({\bf r},t)$ is the local body force acting on the fluid and $p = p({\bf r},t)$ the local pressure at the position ${\bf r} = (x,y,z)$ at time $t$.
$\rho$ and $\nu$ are the mass density and the kinematic viscosity of the fluid, respectively.
The mass transfer of the solute from the capsule to the surrounding medium is governed by the advection-diffusion equation,
\begin{equation}
\frac{\partial c}{\partial t} + {\bf u} \cdot \nabla c = \nabla \cdot \left(D \nabla c\right),
\label{eq:conv_diff}
\end{equation}
where $c = c({\bf r},t)$ is the instantaneous solute concentration and $D = D({\bf r})$ the local diffusion coefficient.
All these equations are solved in both the external medium $\Omega_{\rm out}$, as well as inside the capsule $\Omega_{\rm in}$.
In the following, the inner and the outer fluids are assumed to have the same physical properties, and, thus, $\nu = \nu_{\rm in} = \nu_{\rm out}$, $\rho = \rho_{\rm in} = \rho_{\rm out}$ and $D = D_{\rm in} = D_{\rm out}$.
In the present article, the solutions of these equations are computed with the lattice Boltzmann method (LBM), see the Appendix for more technical details.
\subsection{Initial conditions}
The initial condition for the mass transfer part is,
\begin{equation}
c({\bf r},t=0)=\begin{cases}
c_0, & \text{if ${\bf r} \in \Omega_{\rm in}$}\\
0, & \text{if ${\bf r} \in \Omega_{\rm out}$}
\end{cases}
\text{.}
\label{eq:CI}
\end{equation}
This condition models a uniformly loaded capsule immersed in a solute-free fluid.
It establishes a gradient that triggers mass transfer of the solute from the capsule to the external fluid.
In this study, the surface of the capsule is not maintained at either constant concentration or mass flux in contrast to previous studies.
Thus, the concentration within the sphere decays over time until reaching equilibrium.
For the flow part, the velocity field is initialized with the linear shear flow in the whole domain,
\begin{equation}
{\bf u}({\bf r},t=0) = \left(\gamma z,0,0\right),
\end{equation}
where $\gamma = 2U/H$ is the shear rate.
\subsection{Boundary conditions}
The two parallel plates that bound the domain in the $z$-direction (see Fig.~\ref{fig:numerical_setup}) move in opposite directions along the $x$-axis at constant velocity $U$,
\begin{equation}
{\bf u}(x,y,\pm H/2) = (\pm U,0,0).
\end{equation}
No-slip boundary conditions and impermeability to fluid are set on the membrane of the capsule,
\begin{equation}
{\bf u}_{\rm m} = {\bf u}_{\rm in} = {\bf u}_{\rm out}~~~\text{on}~~~ \partial \Omega_{\rm in},
\end{equation}
where ${\bf u}_{\rm m}$ is the velocity of the capsule membrane.
${\bf u}_{\rm in}$ and ${\bf u}_{\rm out}$ are respectively the velocities of the fluid adjacent to the inner and outer sides of the capsule membrane.
As the capsule center of mass is located in the middle of the computational domain, the capsule membrane  rotates with respect to the $y$-axis without translating.
Periodic velocity boundary conditions are set at the domain edges $x = \pm L/2$ and $y = \pm W/2$.
For the mass transfer part, the boundaries of the domain are maintained at constant concentration $c_\infty = 0$ to model sink condition that corresponds to placing a small particle in a large domain.
Moreover, continuity of both the concentration and the mass flux emerges at the capsule membrane in absence of any interfacial resistance,
\begin{equation}
c(R_{\rm in},t) = c(R_{\rm out},t)~~\text{and}~~
D_{\rm in}\left.\frac{\partial c}{\partial \bf{n}}\right|_{R_{\rm in}} = D_{\rm out}\left.\frac{\partial c}{\partial \bf{n}}\right|_{R_{\rm out}},
\label{eq:continous_BC}
\end{equation}
where $R_{\rm in}$ and $R_{\rm out}$ refer respectively to the inner and the outer sides of the membrane.
These continuous boundary conditions are unsteady and lead to different mass transfer scenarios compared to the largely used constant Dirichlet boundary condition.
They enable to model solute release, in contrast to assuming constant concentration and solving the concentration field only in the external medium.
\subsection{Key physical quantities}
The present problem is expected to be governed by two key dimensionless parameters, the particle-based Reynolds number $\mathrm{Re}$,
\begin{equation}
\mathrm{Re} = \frac{\gamma R^2}{\nu},
\end{equation}
and the Schmidt number $\mathrm{Sc}$,
\begin{equation}
\mathrm{Sc} = \frac{\nu}{D}.
\end{equation}
The explored range of the Reynolds number is $0.01 \leq \mathrm{Re} \leq 1$, while the Schmidt number is hold constant at $\mathrm{Sc} = 10$.
$\mathrm{Re}$ and $\mathrm{Sc}$ are related via the Péclet number ${\rm Pe = Re Sc} = \frac{\gamma R^2}{D}$ that measures the relative importance of advection and diffusion.
The time $t$ is scaled and expressed as ${\rm T} = Dt/R^2$.

Three local observable quantities are measured at the surface of the capsule to characterize and analyze the mass transfer.
The instantaneous surface concentration $c_{\rm s}(\theta,\phi,t)$, which is evaluated using trilinear interpolation of the concentration computed with the LBM at on-lattice grid points.
The variables $\theta$ and $\phi$ refer respectively to the colatitute and the longitude of the spherical coordinates on the capsule surface.
From $c_{\rm s}(\theta,\phi,t)$, one can compute the mass flux at the surface of the capsule as,
\begin{equation}
\varphi(\theta,\phi,t) = \left.-D\frac{\partial c}{\partial {\bf n}}\right|_{r= R_{\rm out}}.
\label{eq:surface_flux}
\end{equation}
The local Sherwood number that is the dimensionless mass transfer coefficient is defined as,
\begin{equation}
{\rm Sh}(\theta,\phi,t) = \frac{2R}{D}\left[\frac{\varphi(\theta,\phi,t)}{c_{\rm s}(\theta,\phi,t) - c_\infty}\right].
\label{eq:surface_Sh}
\end{equation}
The instantaneous average Sherwood number is computed by integration of ${\rm Sh}\left(\theta,\phi,t\right)$ over the surface of the capsule $\partial \Omega_{\rm in}$,
\begin{equation}
{\rm Sh}(t) = \frac{1}{4\pi R^2}\int_{\partial \Omega_{\rm in}} {\rm Sh}(\theta,\phi,t){\rm d}S.
\label{eq:average_Sh}
\end{equation}
The present mathematical formulation holds also for heat transfer problem, by replacing the concentration by the temperature, the solute diffusion coefficient by the thermal diffusivity, the Schmidt number by the Prandtl number, and the Sherwood number by the Nusselt number.
The initial and boundary conditions studied in this article correspond to cooling a hot spherical particle by a cold sheared fluid.
\begin{figure*}
\centering
\includegraphics[width=0.8\textwidth]{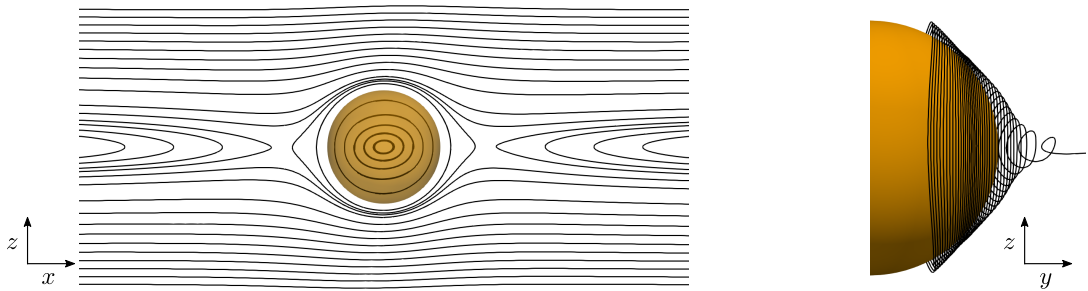}
\caption{Induced flow streamlines inside and around the capsule under applied shear flow at ${\rm Re} = 1$. Left: streamlines in the $x$-$z$ plane passing by the center of the capsule. Right: streamlines in the $y$-$z$ plane exhibiting the spiraling nature of the flow. Similar type of streamlines topology is obtained within the range $0.01 \leq {\rm Re} \leq 1$.}
\label{fig:flow_pattern}
\end{figure*}
\section{Results and discussions}
\label{sec:results_discussions}
The effect of flow on solute mass transfer from a spherical capsule freely suspended in a sheared fluid is studied here.
The Reynolds number is varied within the range $0.01 \leq {\rm Re} \leq 1$ by varying solely the wall velocity $U$, while keeping all the other parameters constant.
The Schmidt number is set to ${\rm Sc = 10}$ so that the Péclet number varies from $0.1$ to $10$ in order to cover both diffusion and advection dominated regimes.
The initial concentration of the solute in the capsule is set to $c_0 = 1$.
\subsection{Fully developed flow pattern}
Figure \ref{fig:flow_pattern} shows the flow streamlines computed at ${\rm Re} = 1$, which are the typical patterns in the range $0.01 \leq {\rm Re} \leq 1$.
The left panel of Fig.~\ref{fig:flow_pattern} depicts the streamlines in the $x$-$z$ plane that passes by the center of the capsule.
The encapsulated fluid flow exhibits closed ellipsoidal lines as the capsule membrane rotates under the applied external shear flow.
There is a region of closed streamlines around the particle, that tend to elongate and adopt an eye-shaped pattern.
Because closed streamlines are known to hinder transport by advection, mass transfer is thus diffusion-dominated in the vicinity of the particle.
The region of closed streamlines is surrounded by a pair of recirculating wakes, originating from inertial forces \cite{Subramanian2006}.
These recirculations are expected to transport the solute far away from the capsule and to enhance mass transfer by forced convection.
The streamlines remain parallel to the channel walls far away from the particle, as in a simple shear flow in absence of any obstacle.
The rotation of the capsule membrane generates a spiraling flow in its neighborhood, as illustrated in the right panel of Fig.~\ref{fig:flow_pattern}.
The streamlines emerge from the surface of the capsule forming spirals up to the far flow field.
The computed streamlines within the present study have similar topology as in Refs.~\cite{Subramanian2006,Haddadi2015,Wang2019}.
\subsection{Effect of flow on solute spatial distribution}
\begin{figure*}
\centering
\includegraphics[width=0.4\textwidth]{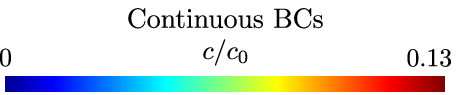}
\hspace{1.5cm}
\includegraphics[width=0.4\textwidth]{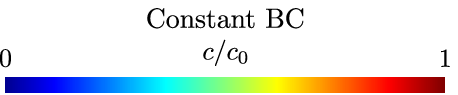}
\\[0.2cm]
\includegraphics[width=0.4\textwidth]{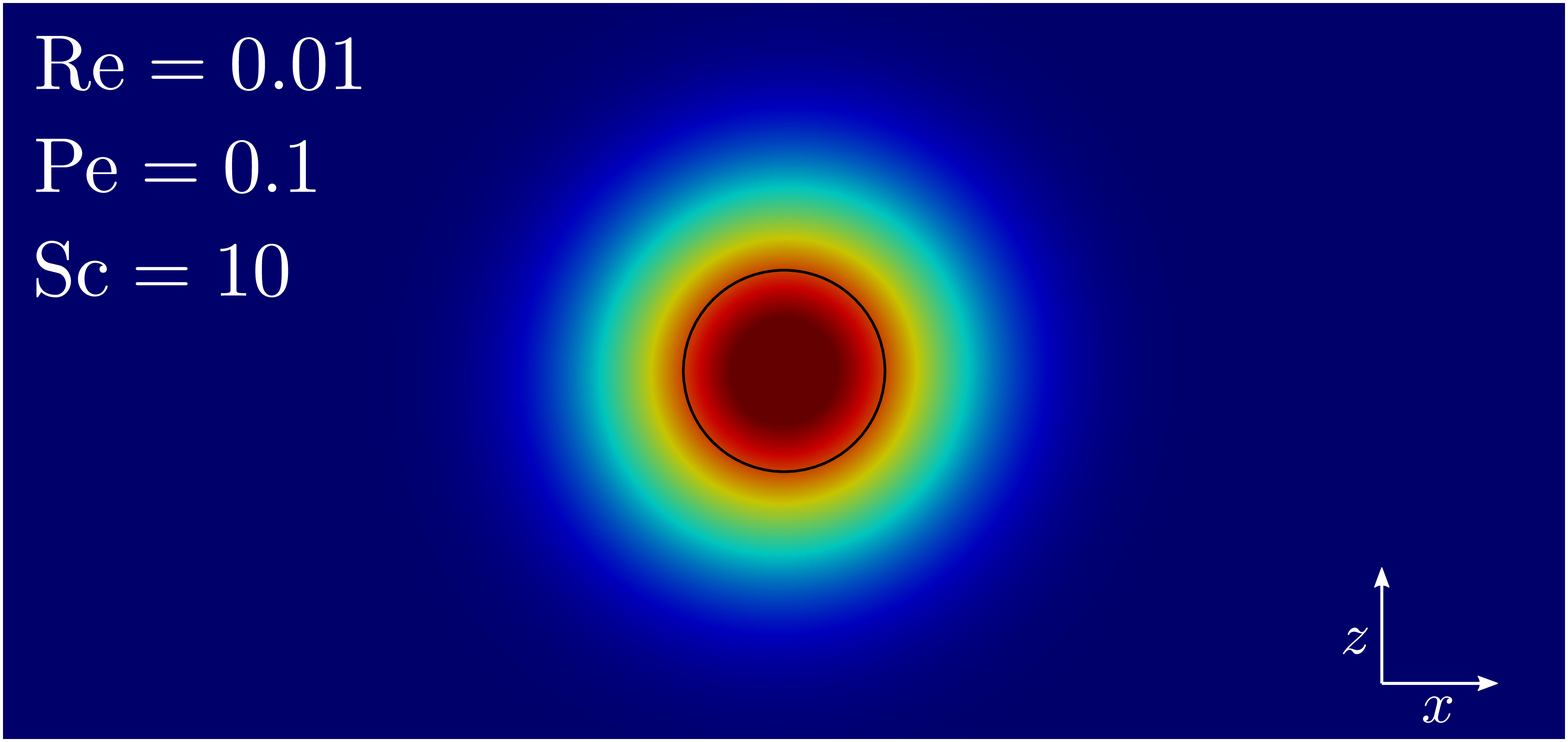}
\hspace{1.5cm}
\includegraphics[width=0.4\textwidth]{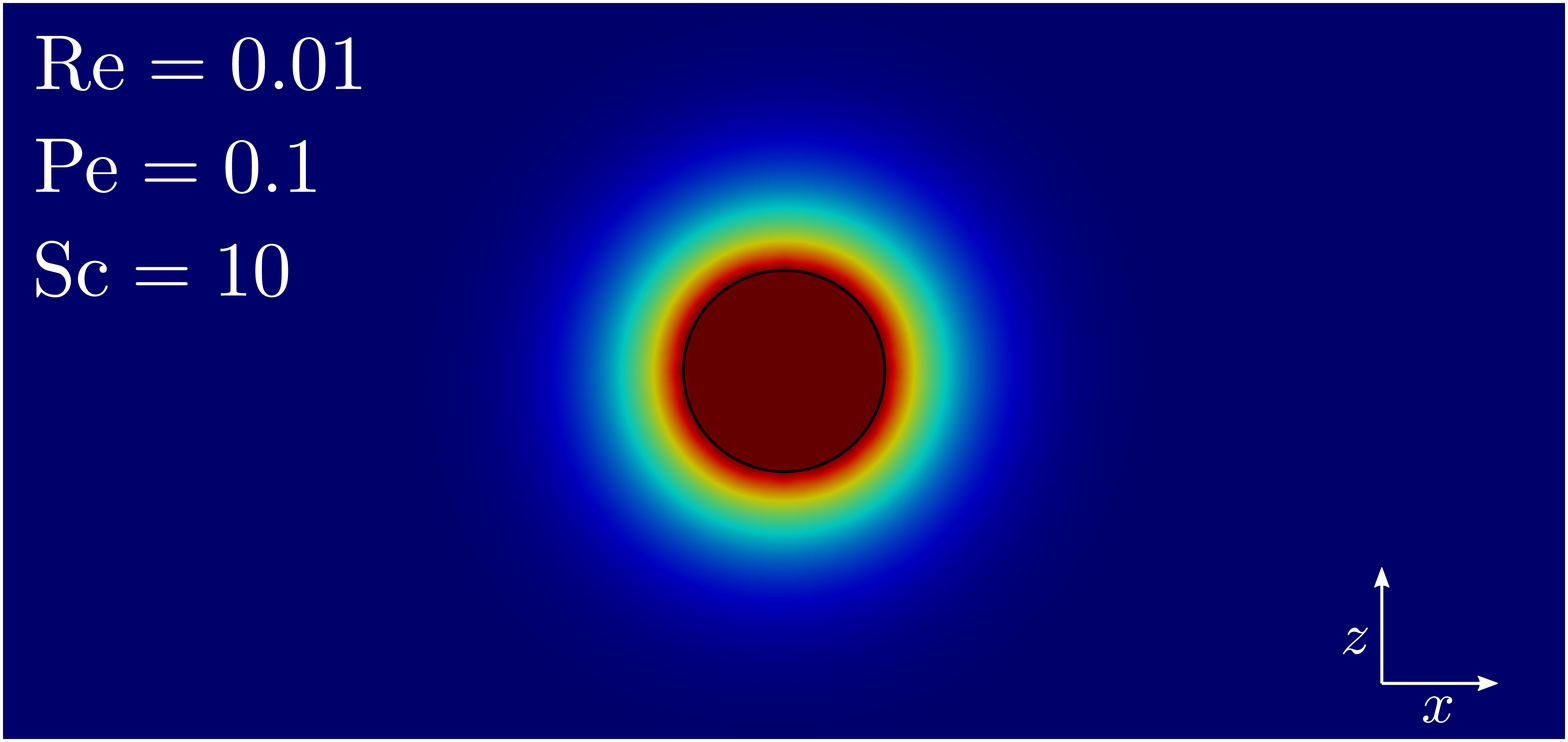}
\\[0.2cm]
\includegraphics[width=0.4\textwidth]{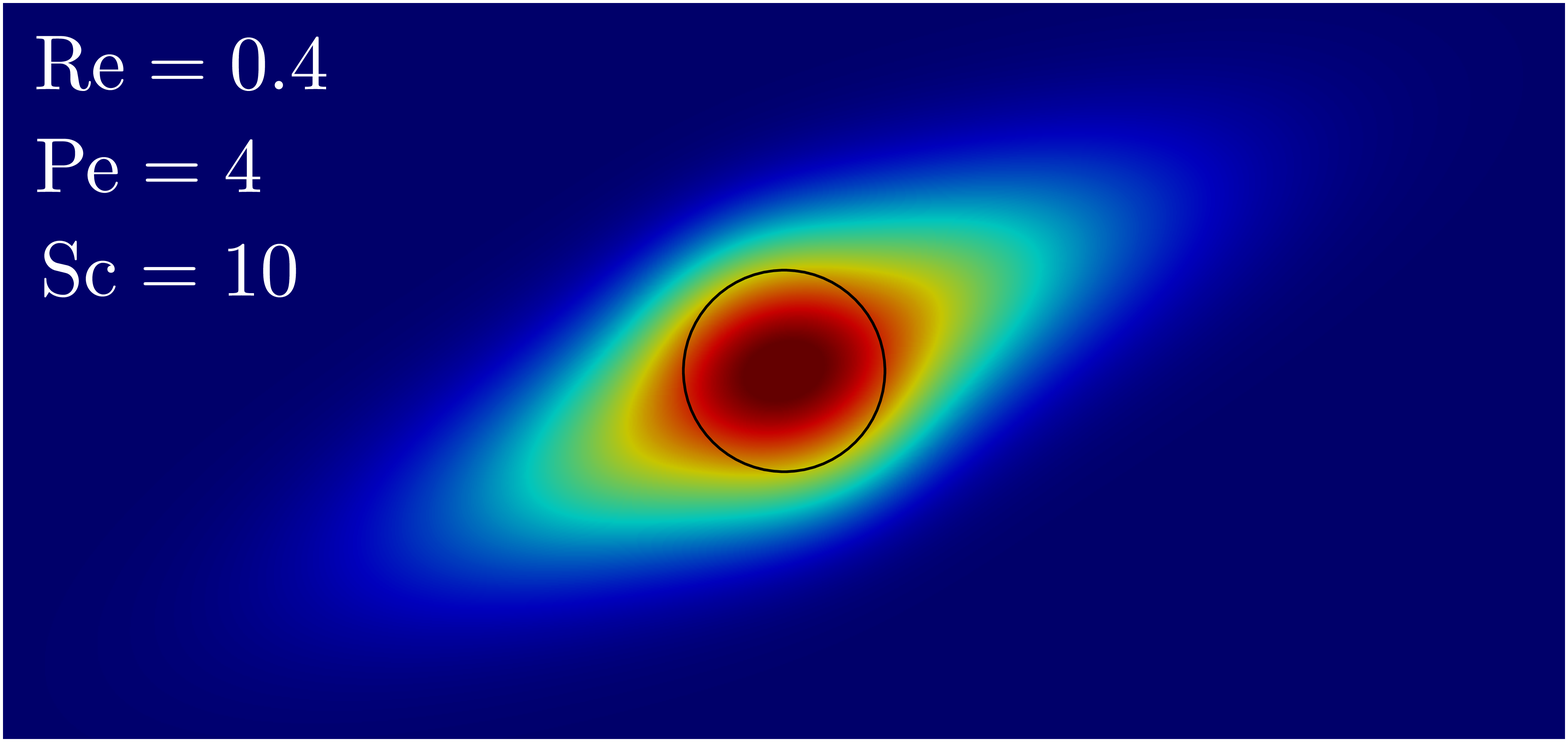}
\hspace{1.5cm}
\includegraphics[width=0.4\textwidth]{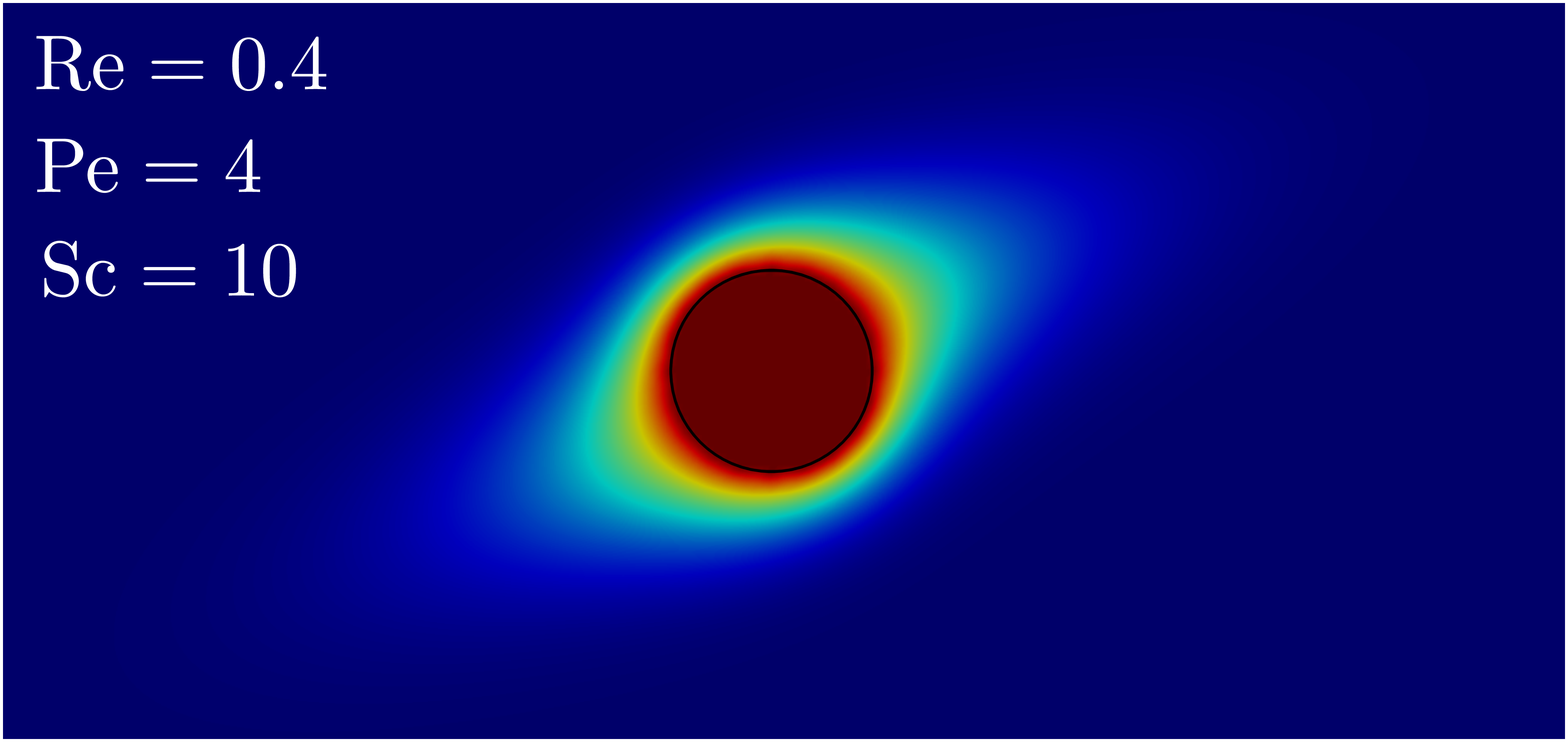}
\\[0.2cm]
\includegraphics[width=0.4\textwidth]{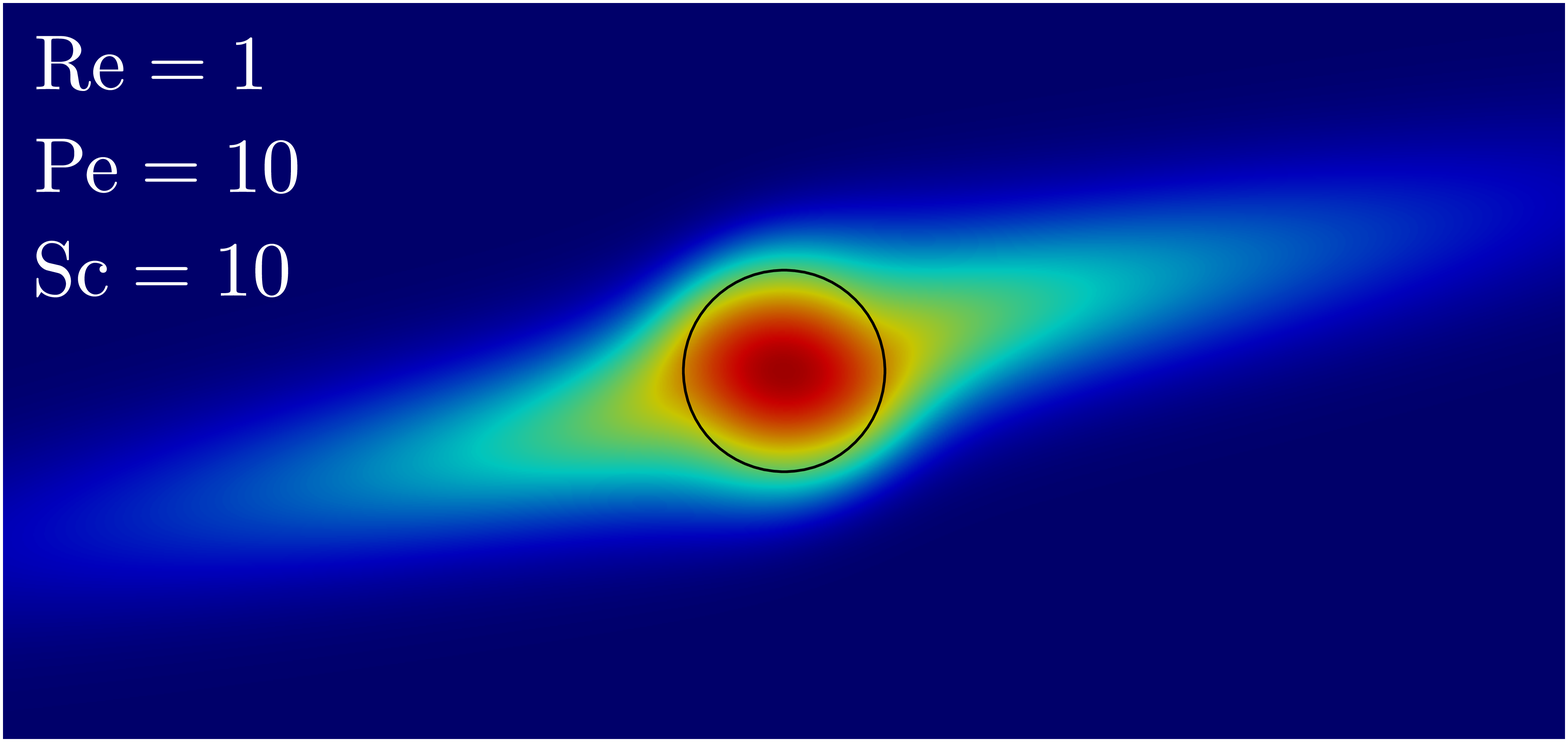}
\hspace{1.5cm}
\includegraphics[width=0.4\textwidth]{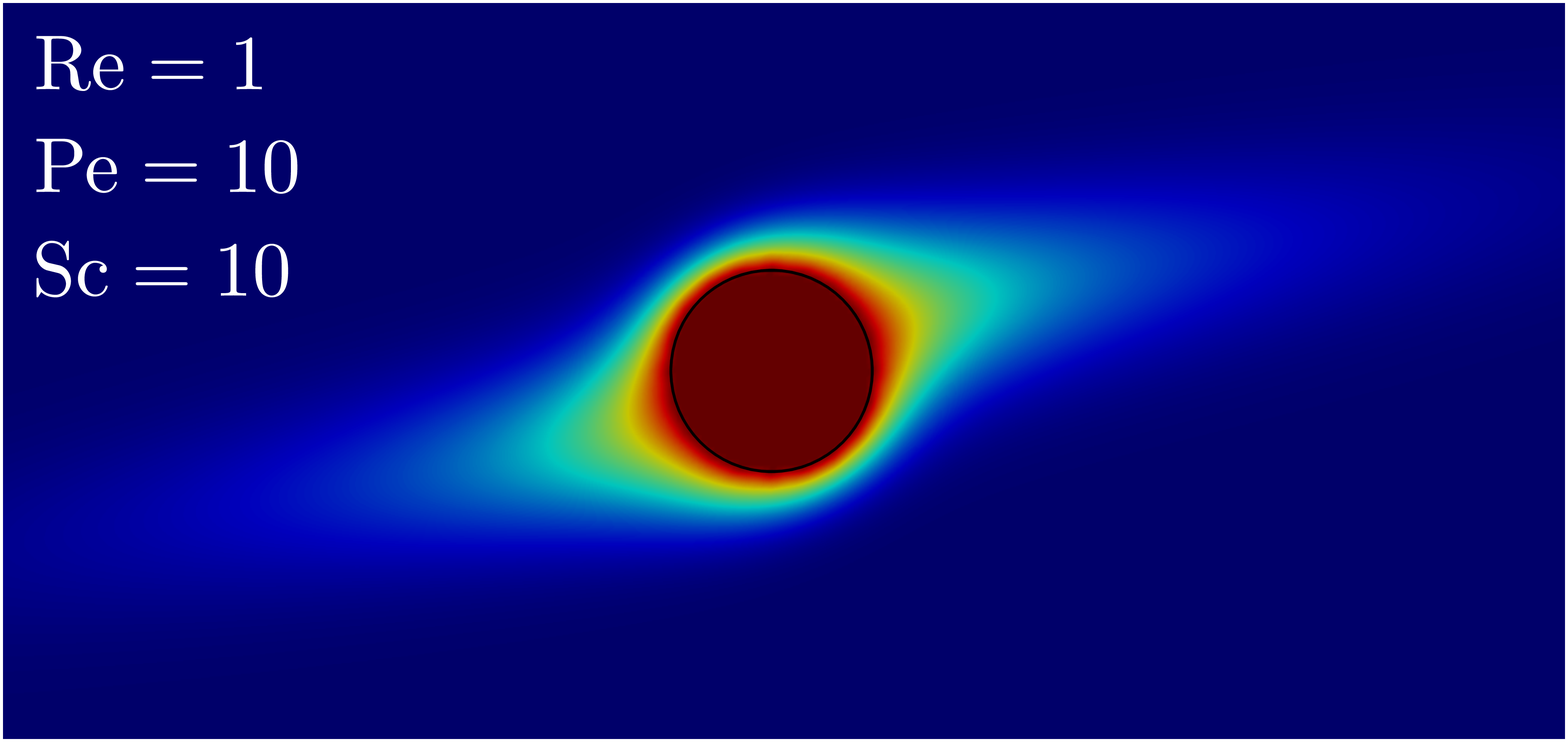}
\caption{Snapshots of solute concentration fields computed at dimensionless time ${\rm T} = 0.65$ under various Reynolds numbers ${\rm Re} = 0.01$, $0.4$ and $1$ when considering continuous boundary conditions (left column) and constant Dirichlet boundary condition (right column) at the surface of the capsule. The Schmidt number is ${\rm Sc} = 10$. The Péclet number ${\rm Pe}$ varies from $0.1$ to $10$; and thus, it covers both diffusion and advection dominated regimes.}
\label{fig:concentration_fields}
\end{figure*}

The effect of flow on the solute spatial distribution is shown in Fig.~\ref{fig:concentration_fields} by reporting the concentration isocontours at various Reynolds numbers ${\rm Re} = 0.01$, $0.4$ and $1$ computed at the same dimensionless time ${\rm T} = 0.65$.
For comparison purposes, simulations performed with constant concentration at the capsule surface, as extensively considered in the literature, are also shown (right column).
Qualitatively the obtained concentration distribution is almost similar for both boundary conditions; however, the overall concentration is higher for the case of Dirichlet boundary condition that sustains the initial concentration to $c_0=1$.
For ${\rm Re = 0.01}$, mass transfer is dominated by diffusion.
Thus, the resulting concentration contours are radial and centered with respect to the capsule center of mass.
At higher Reynolds numbers, advection is the dominant mass transfer mechanism.
The solute is transported efficiently by the flow far away from the capsule.
It exhibits stretched concentration isocontours along the elongational direction of the shear flow.
This direction inclines towards the channel centerline as the Reynolds number is increased since the solute has less time to diffuse in the transverse direction.
The concentration is not uniform inside the capsule when considering continuous boundary conditions on the membrane (left column), compared to the case of Dirichlet boundary conditions (right column).
It has maximum at the center and it decays as it approaches the membrane.
\subsection{Surface mass transfer quantities}
\begin{figure*}
\centering
\includegraphics[width = 5.5cm, trim={4.6cm 1cm 4.6cm 0cm}, clip]{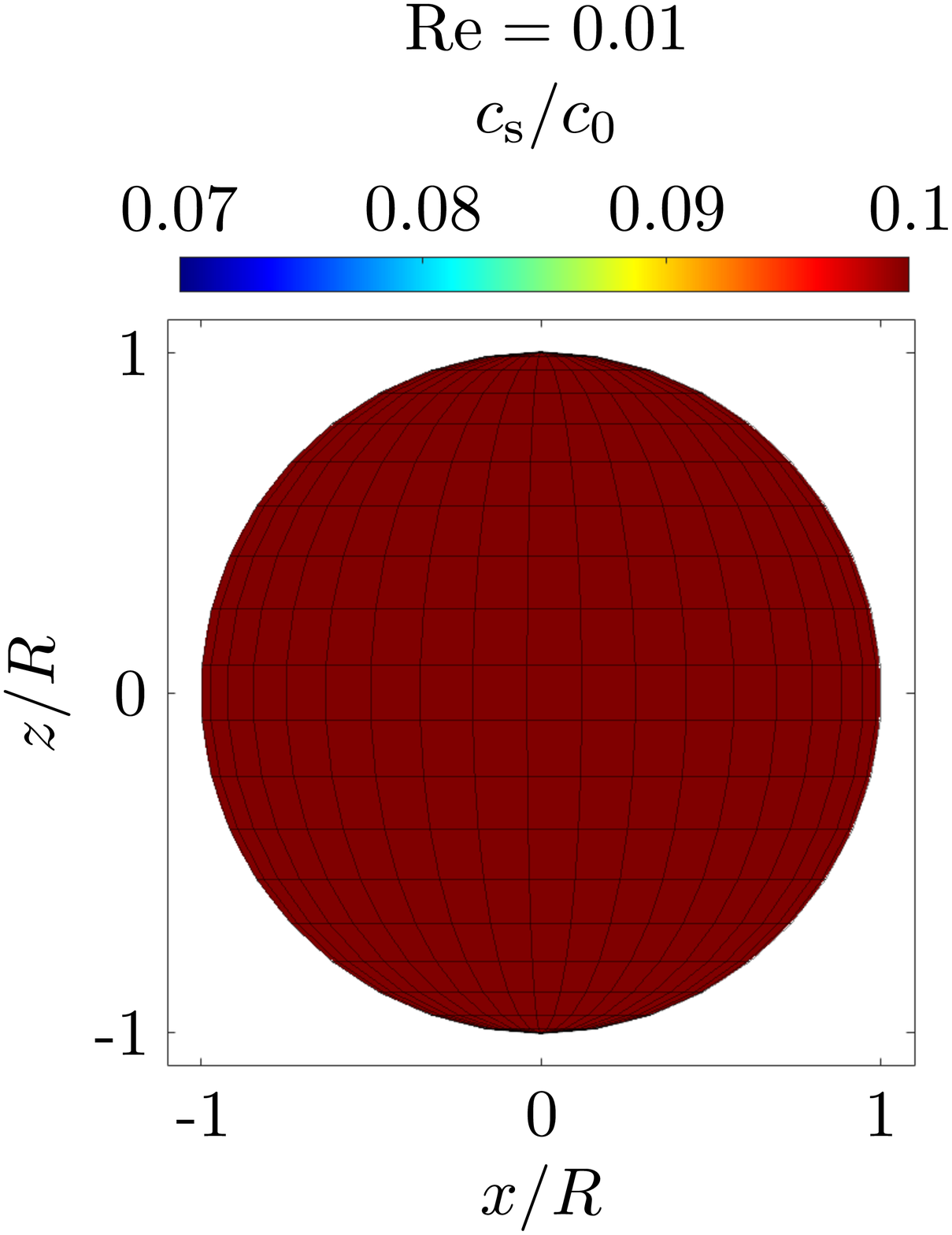}
\includegraphics[width = 5.5cm, trim={4.6cm 1cm 4.6cm 0cm}, clip]{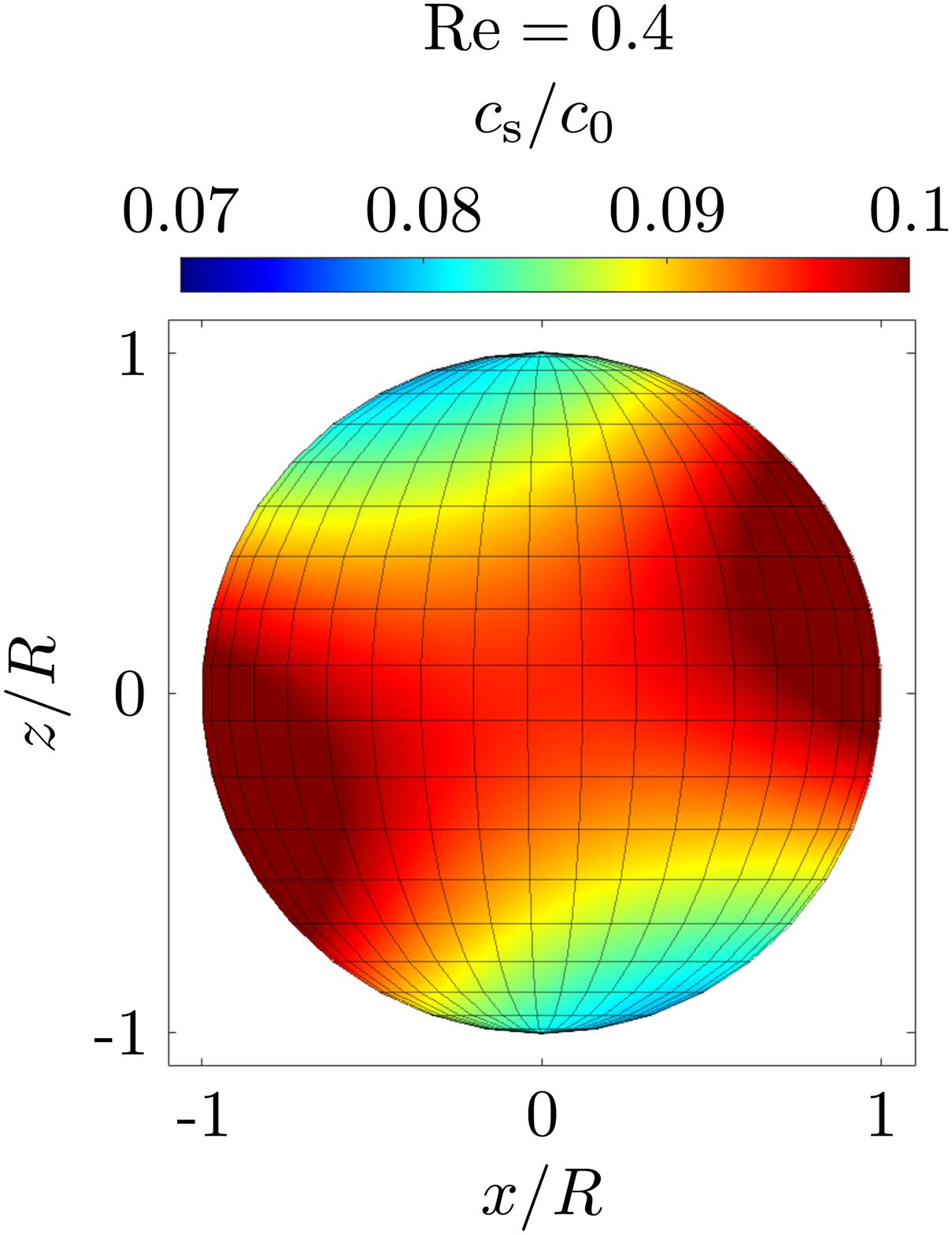}
\includegraphics[width = 5.5cm, trim={4.6cm 1cm 4.6cm 0cm}, clip]{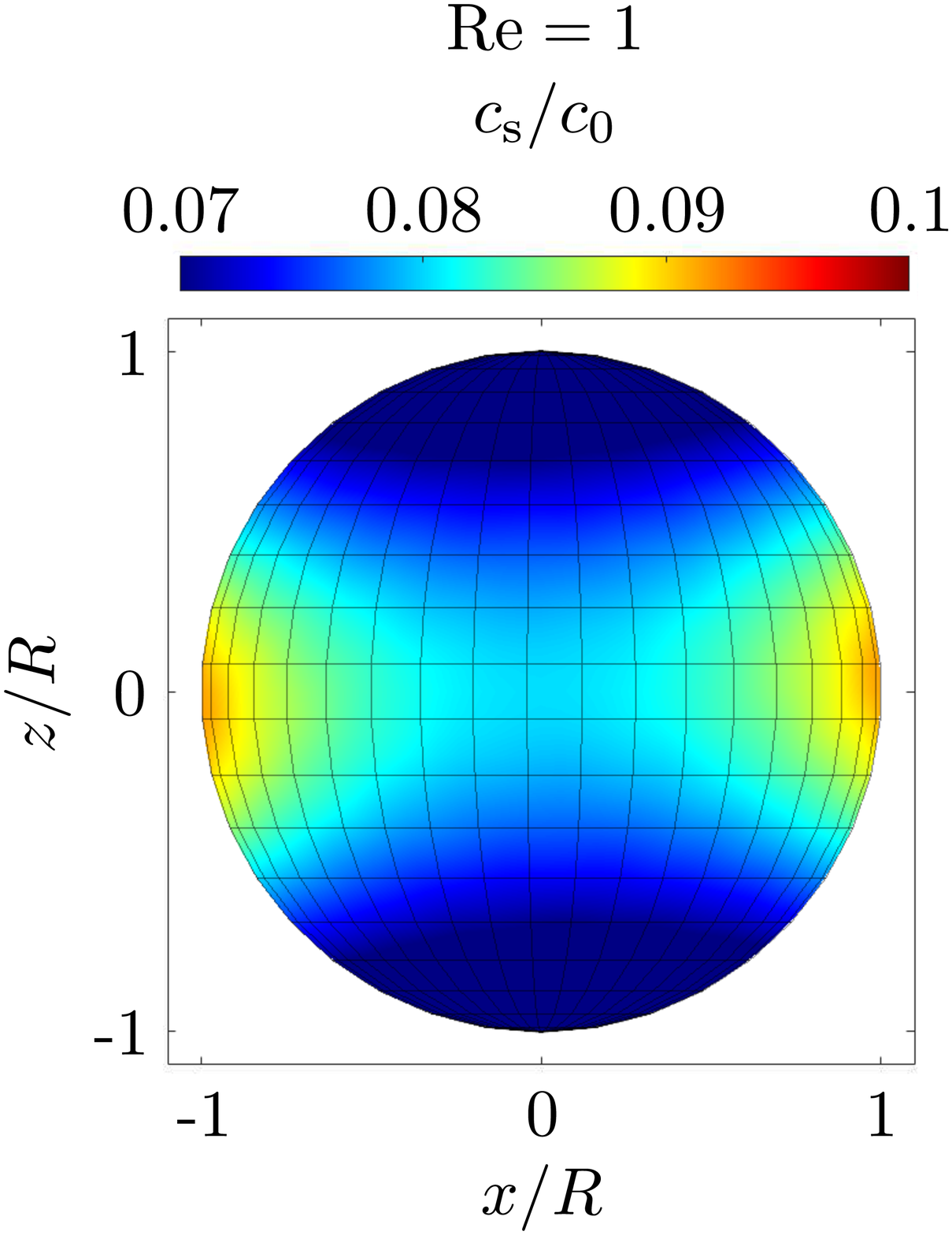}\\[0.3cm]
\includegraphics[width = 5.5cm, trim={4.6cm 1cm 4.6cm 0cm}, clip]{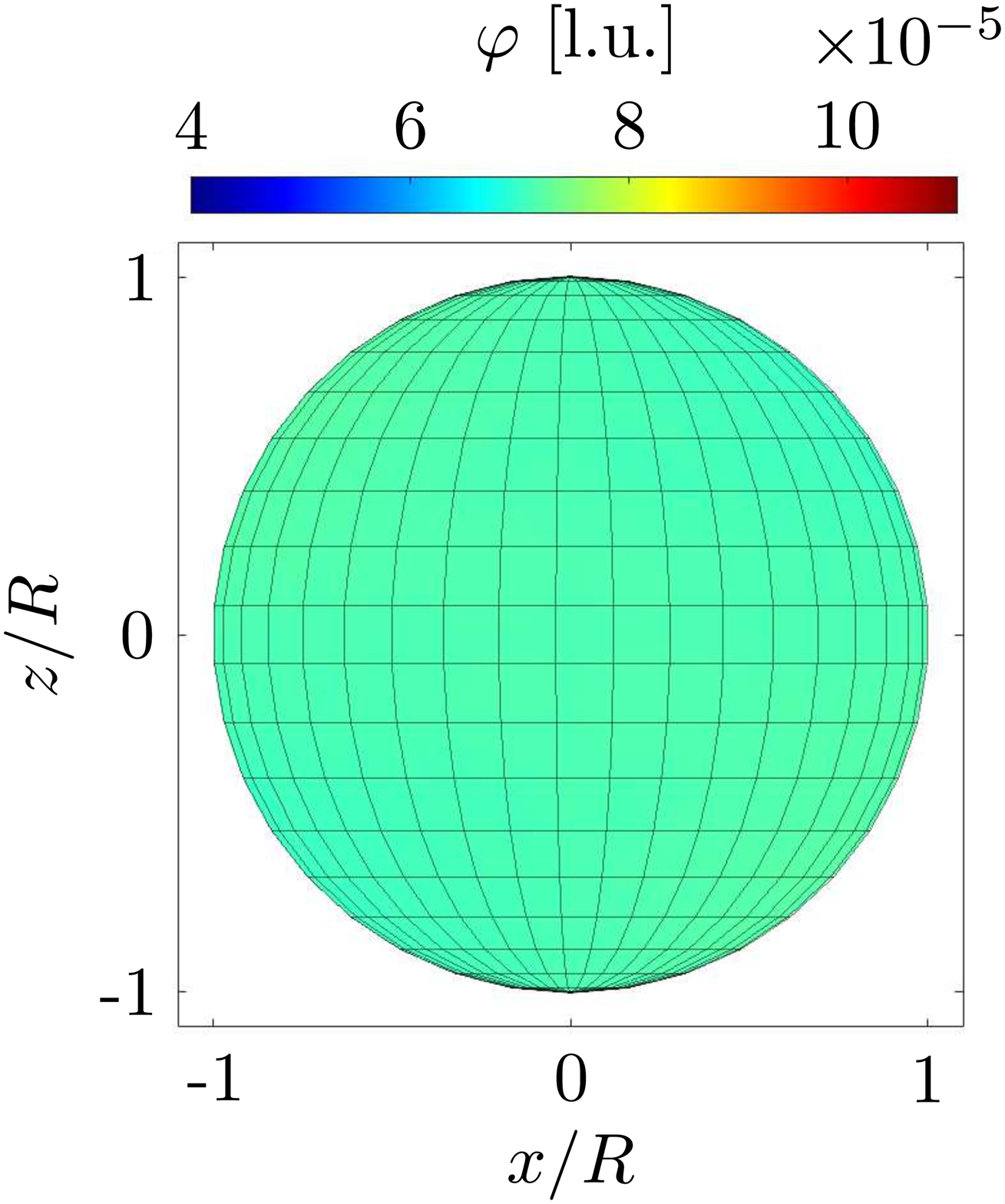}
\includegraphics[width = 5.5cm, trim={4.6cm 1cm 4.6cm 0cm}, clip]{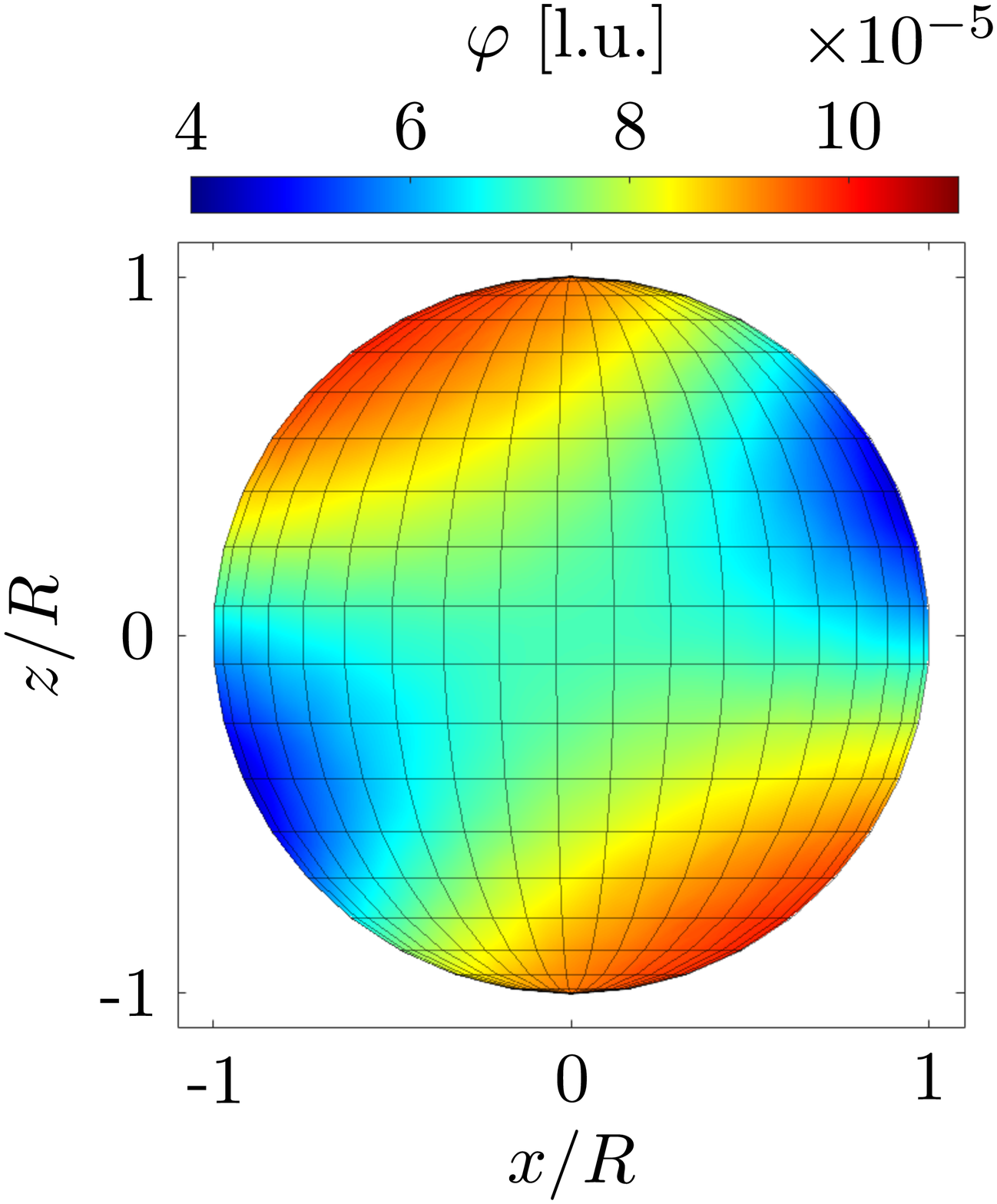}
\includegraphics[width = 5.5cm, trim={4.6cm 1cm 4.6cm 0cm}, clip]{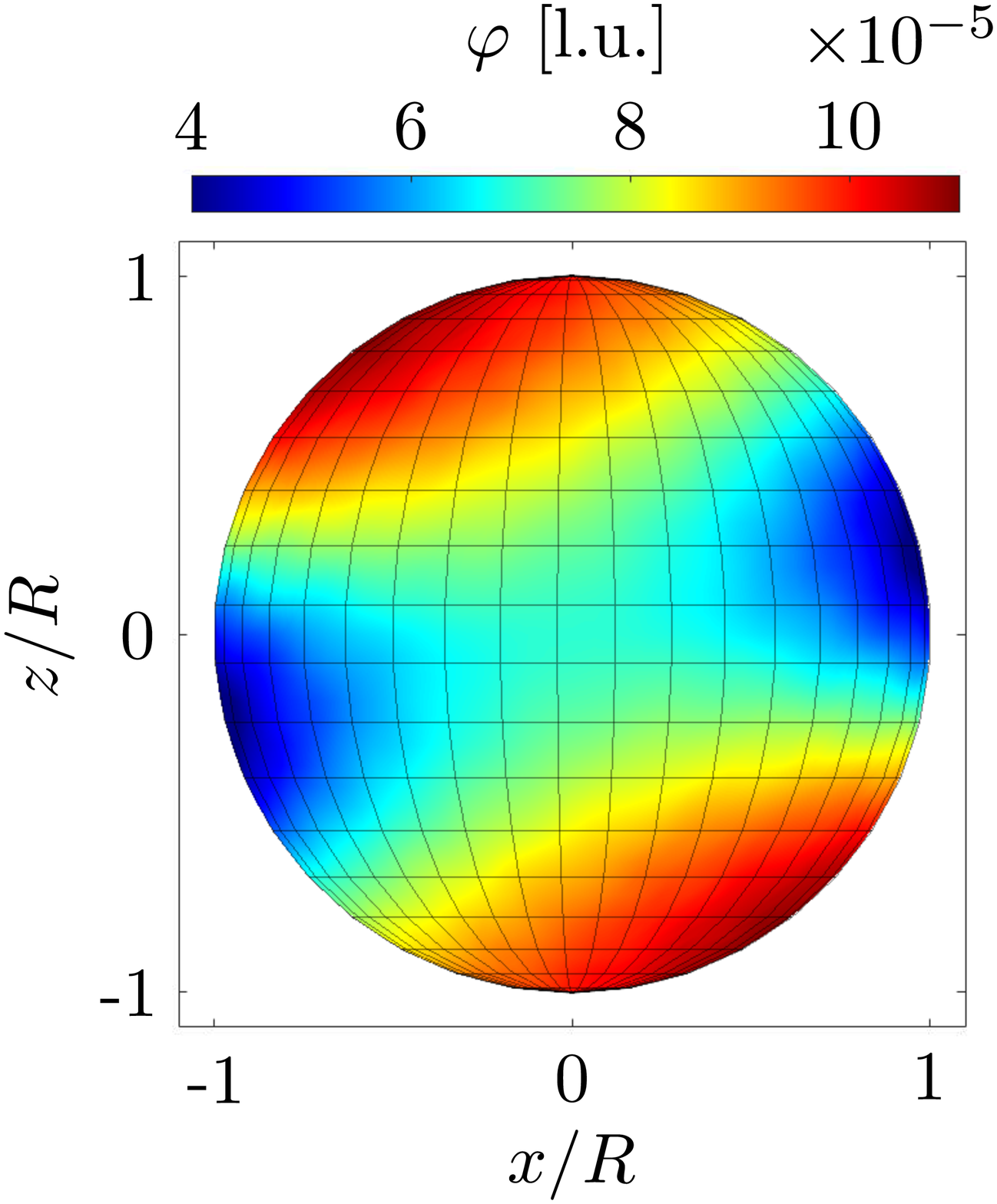}\\[0.3cm]
\includegraphics[width = 5.5cm, trim={4.6cm 1cm 4.6cm 0cm}, clip]{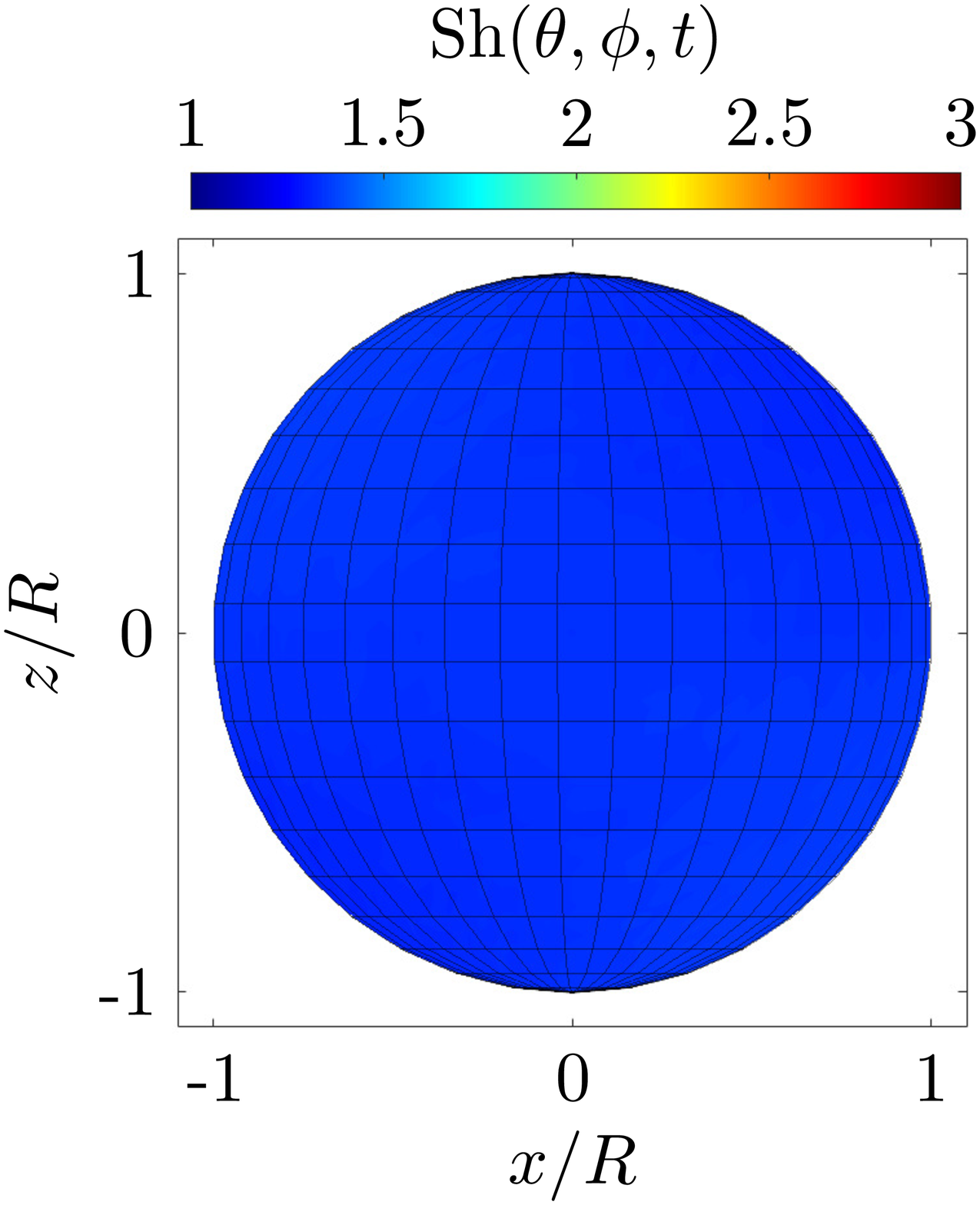}
\includegraphics[width = 5.5cm, trim={4.6cm 1cm 4.6cm 0cm}, clip]{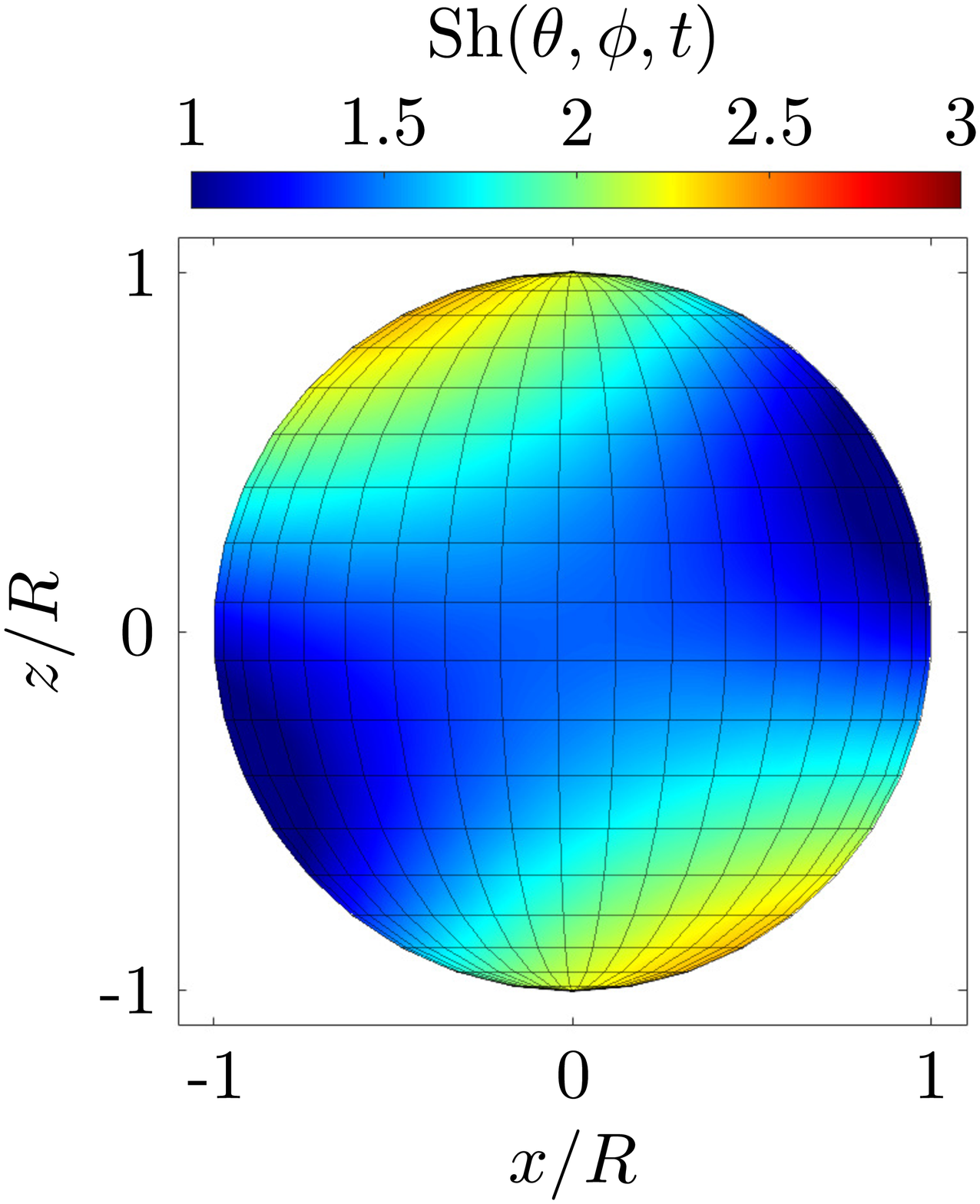}
\includegraphics[width = 5.5cm, trim={4.6cm 1cm 4.6cm 0cm}, clip]{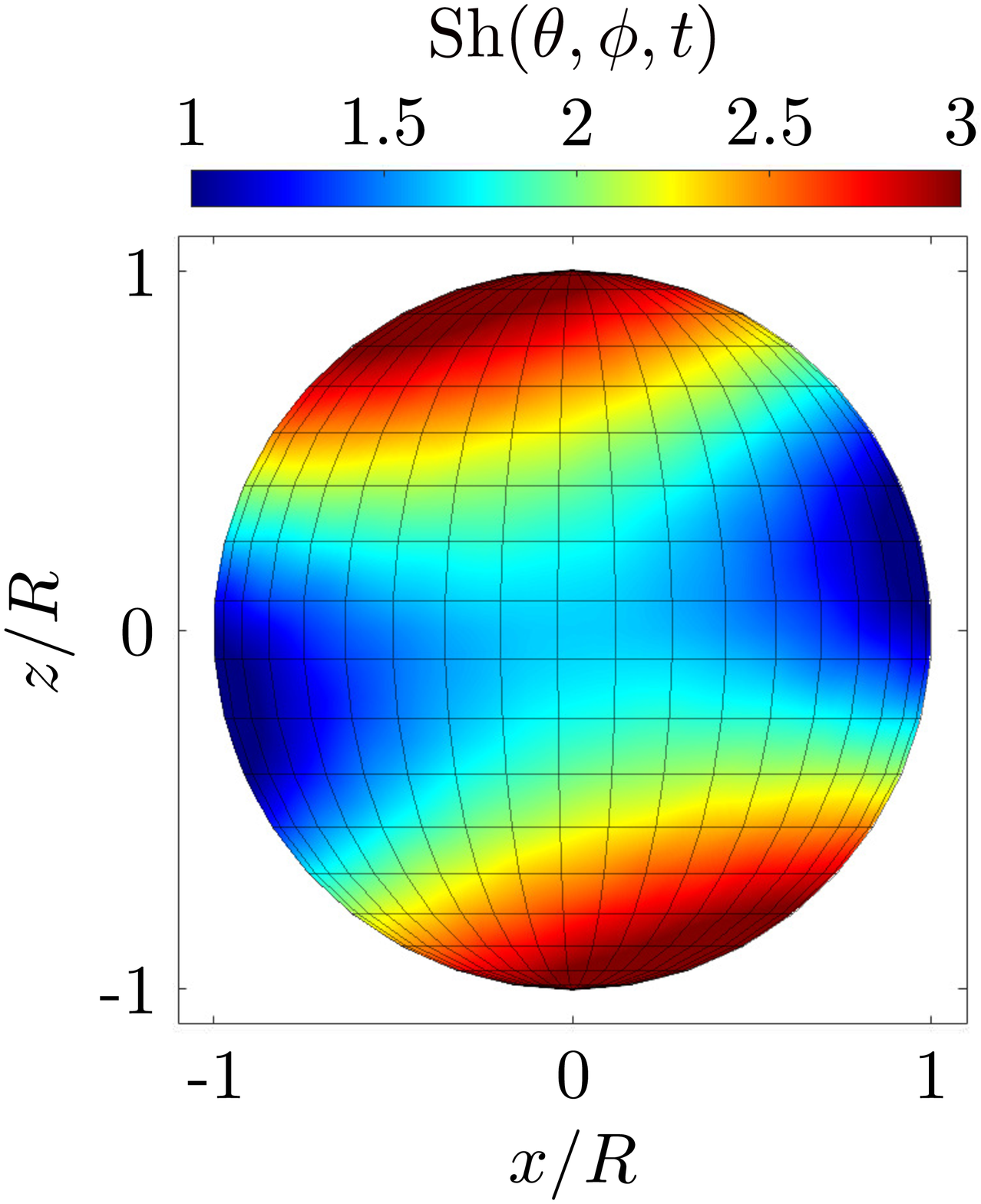}
\caption{Local mass transfer quantities computed on the capsule surface at dimensionless time ${\rm T} = 0.65$ for various Reynolds numbers ${\rm Re} = 0.01$ (left column), ${\rm Re} = 0.4$ (middle column) and ${\rm Re} = 1$ (right column). From top to bottom: normalized surface concentration $c_{\rm s}(\theta,\phi,t)/c_0$, surface mass flux $\varphi(\theta,\phi,t)$ and local Sherwood number ${\rm Sh}(\theta,\phi,t)$. The shear flow enhances mass transfer in some regions on the capsule, which renders these quantities nonuniform. The Schmidt number is ${\rm Sc} = 10$. [l.u.]: LBM lattice units.}
\label{fig:local_quantities}
\end{figure*}

The concentration $c_{\rm s}(\theta,\phi,t)$, the mass flux $\varphi(\theta,\phi,t)$ and the Sherwood number ${\rm Sh}(\theta,\phi,t)$ at the surface of the capsule measured at dimensionless time ${\rm T} = 0.65$ are plotted in Fig.~\ref{fig:local_quantities} for various Reynolds numbers ${\rm Re = 0.01}$ (left column), $0.4$ (middle column) and $1$ (right column).
At low Reynolds numbers, for example at ${\rm Re} = 0.01$, these quantities are uniform all over the surface of the capsule because the solute mass transfer takes place mainly by diffusion.
At high Reynolds numbers, the contribution of advection to mass transfer becomes important and the resulting local surface quantities show strong spatial variations.
The surface concentration is maximal in the two dark red opposite areas, where the solvent pressure is low.
There, mass transfer is dominated by diffusion and the released solute is weakly transported by convection.
Thus, the solute accumulates and remains in these regions resulting in weak mass flux and Sherwood number.
The areas of maximal concentration inclines towards the channel centerline ($z = 0$) as the Reynolds number is increased.
On the contrary, the concentration is minimal on the two diametrically opposed surfaces that are the most exposed to the flow (dark blue).
They are located at approximately $90$ degrees from the regions of maximal concentration.
In these regions, advection prevails and the newly released solute is efficiently transported by the flow.
The mass flux is, thus, particularly high there resulting in a maximal Sherwood number.
The shear flow enhances mass transfer in these regions on the capsule, which renders all the mass transfer quantities anti-symmetric with respect to the plane of zero velocity.
\subsection{Steady Sherwood number}
\begin{figure*}[t]
\centering
\includegraphics[width = 0.6\textwidth]{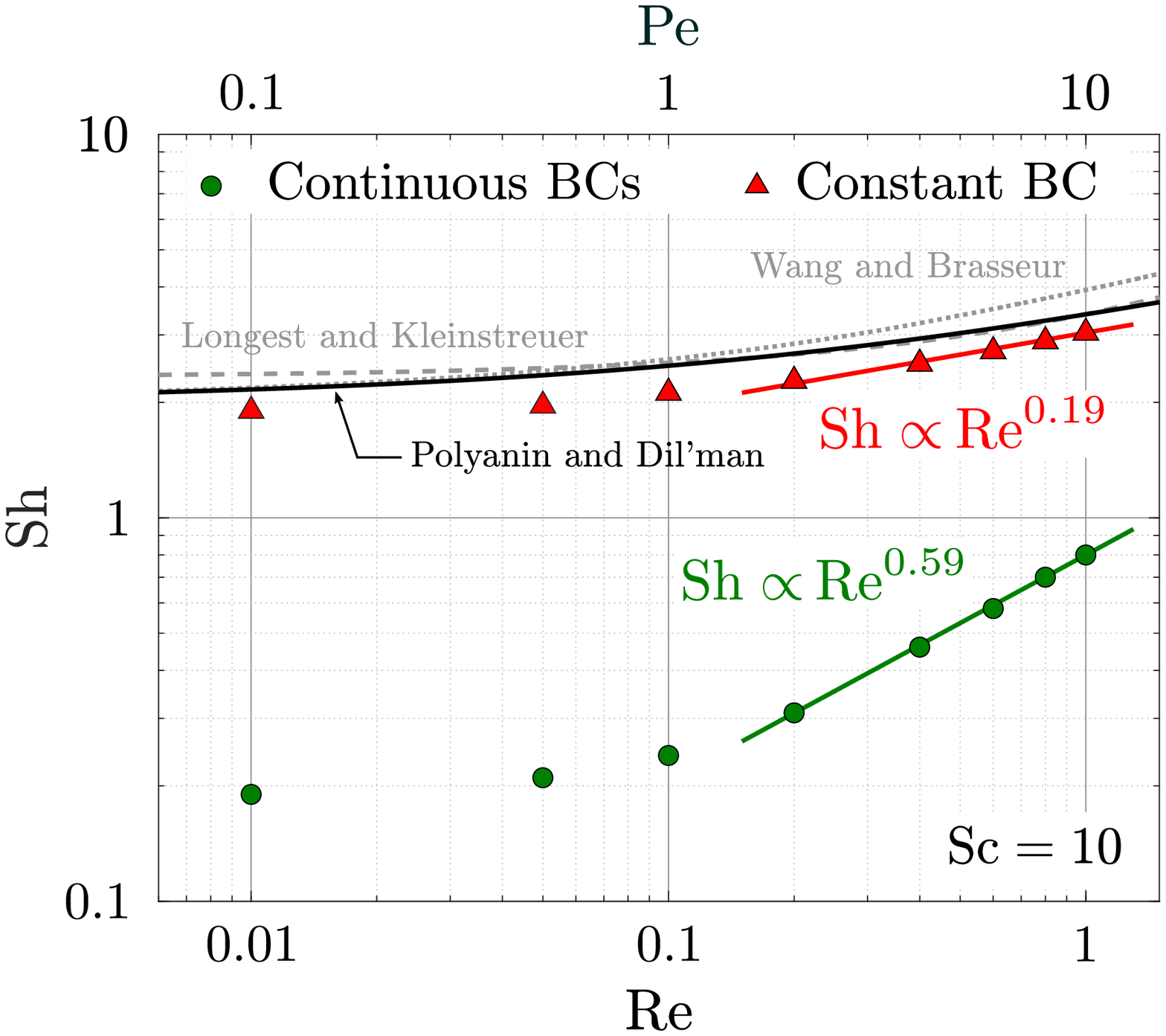}
\caption{The steady average Sherwood number ${\rm Sh}$ \textit{vs.} the Reynolds number ${\rm Re}$ and the Péclet number ${\rm Pe}$ computed at a Schmidt number ${\rm Sc} = 10$, while considering either unsteady continuous (circles) or constant Dirichlet (triangles) boundary conditions at the surface of a spherical capsule rotating under shear flow. The correlations of Wang and Brasseur \cite{Wang2019}, Polyanin and Dil'man \cite{Polyanin1985}, and Longest and Kleinstreuer \cite{Longest2004} are obtained for constant boundary condition and are shown for comparison purposes. ${\rm Sh}$ is dramatically altered by the type of the boundary conditions set at the particle surface. The figure suggests ${\rm Sh} \propto {\rm Re}^{0.59}$ for continuous boundary conditions and ${\rm Sh} \propto {\rm Re}^{0.19}$ for Dirichlet boundary condition in the advection-dominated regime (${\rm Pe} > 1$).}
\label{fig:steady_Sh}
\end{figure*}
The instantaneous average Sherwood number ${\rm Sh(t)}$, Eq.~\eqref{eq:average_Sh}, reaches a steady value ${\rm Sh}$ as the concentration boundary layer adopts a steady thickness around the capsule. 
${\rm Sh}$ is reported in Fig.~\ref{fig:steady_Sh} as a function of the Reynolds and the Péclet numbers in logarithmic scale with $\rm Sc = 10$.
The steady value increases with the Reynolds number since the mass transfer is enhanced by forced convection and local shear rate in the neighborhood of the particle.
For comparison purposes, the values of ${\rm Sh}$ computed when considering constant concentration at the surface of the capsule are also represented, along with the correlations of Wang and Brasseur \cite{Wang2019}, Polyanin and Dil'man \cite{Polyanin1985}, and Longest and Kleinstreuer \cite{Longest2004}, which are given by,
\begin{itemize}
\item Correlation of Wang and Brasseur \cite{Wang2019}:
\begin{equation}
{\rm Sh} = {\rm Sh_0} + 0.03748{\rm Pe}^{0.674}{\rm Re}^{0.583 - 0.032\ln {\rm Pe}},
\label{eq:Sh_Wang_Brasseur1}
\end{equation}
where ${\rm Sh_0}$ is the Sherwood number in the limit of ${\rm Re}\rightarrow 0$,
\begin{equation}
{\rm Sh_0}=\begin{cases}
2+0.580{\rm Pe}^{0.5} & {\rm Pe} \leq 5\\
2.438{\rm Pe}^{0.187} & 5 \leq {\rm Pe} \leq 100 \\
9 - 16.128{\rm Pe}^{-0.349} & {\rm Pe} \geq 100
\label{eq:Sh_0_Wang_Brasseur2}
\end{cases}
\end{equation}
\item Correlation of Polyanin and Dil'man \cite{Polyanin1985}:
\begin{equation}
{\rm Sh} = 2 + \frac{0.52{\rm Pe}^{1/2}}{1 + 0.057{\rm Pe}^{1/2}},~~~~ {\rm Re} \rightarrow 0,~~0 < {\rm Pe} < \infty
\end{equation}
\item  Correlation of Longest and Kleinstreuer \cite{Longest2004}:
\begin{equation}
{\rm Sh} = C_1 + C_2{\rm Re}_d^{C_3}{\rm Sc}^{0.333},
\end{equation}
with,
\begin{align}
& C_1 =  1.92 + 1.03{\rm B},\\
& C_2 =  0.42\exp\left( -2.08{\rm B} \right),\\
& C_3 = 0.53 + 0.47{\rm B},
\end{align}
where ${\rm Re}_d = \frac{\gamma d^2}{\nu}$, $h$ is the half distance between the moving walls, $d$ is the particle diameter and $\mathrm{B} = d/h$ is the blockage ratio.
This correlation is valid for $0 \leq {\rm Re}_d \leq 32$, and $ 0.1 \leq \mathrm{B} < 1$.
\end{itemize}
For both boundary conditions, two distinct behaviors are observed depending on the Péclet number.
When ${\rm Pe} < 1$, mass transfer is dominated by diffusion and the Sherwood number is barely affected by the flow.
It is relatively low and increases slowly with ${\rm Re}$.
For ${\rm Pe} > 1$, advection becomes the dominant mass transfer mechanism and the resulting ${\rm Sh}$ increases linearly in the logarithmic scale.
These two characteristic behaviors have also been reported for other types of particles subjected to various flow conditions, see e.g. Ref.~\cite{Clift1978}.
However, Fig.~\ref{fig:steady_Sh} highlights for the first time how the Sherwood number depends strongly on the boundary conditions set at the surface of a particle.
Because the mass flux is lower when the concentration at the capsule surface is not sustained at its initial high value, the unsteady continuous boundary conditions reduce dramatically ${\rm Sh}$.
But meanwhile, they lead to a stronger dependency of ${\rm Sh}$ on ${\rm Re}$ in the advection-dominated regime (${\rm Pe > 1}$), 
\begin{equation}
{\rm Sh} \propto {\rm Re}^{0.59},
\end{equation}
with an exponent that is greater than ${0.19}$ obtained in the present study when setting constant concentration at the capsule surface, and which is consistent with ${0.187}$ obtained by Wang and Brasseur \cite{Wang2019}.

In the advection-dominated regime, the present numerical data obtained for constant boundary condition are slightly lower than the correlations of Polyanin and Dil'man \cite{Polyanin1985} and Longest and Kleinstreuer \cite{Longest2004}.
But, this deviation is not as significant as the large difference induced by the type of the boundary conditions set at the capsule surface.
For these constant boundary conditions, ${\rm Sh} \rightarrow 2$ as ${\rm Re} \rightarrow 0$, which is the expected value for pure diffusion from a sphere sustained at constant concentration \cite{Clift1978}.
For the unsteady continuous boundary condition, ${\rm Sh}$ tends to a lower value than $2$ as ${\rm Re} \rightarrow 0$.
\section{Conclusions}
\label{sec:conclusion}
A new numerical method is proposed to study unsteady solute release from a spherical rigid capsule subjected to shear flow.
It is based on three-dimensional and two-component lattice Boltzmann method to compute both the flow and the solute advection-diffusion inside as well as outside the capsule, and on the immersed boundary method for the fluid-structure interaction coupling.
The method allows to set unsteady and continuous boundary conditions on the membrane of the capsule, which differ from the constant and uniform surface concentration (Dirichlet boundary condition) or mass flux (Neumann boundary condition) largely considered in the literature.
These boundary conditions are adapted to model solute release from particles, for which the solute concentration inside the particle decays over time.

The effect of the flow on the concentration, mass flux and Sherwood number at the capsule surface are also reported and discussed.
Having continuity of both the concentration and the mass flux at the surface of the capsule leads to significantly lower Sherwood numbers than when the surface is maintained at constant concentration.
Unsteady continuous boundary conditions also result in a stronger dependency of the global Sherwood number on the Reynolds number measured by a larger exponent.
In the advection-dominated regime, ${\rm Sh}$ is found to scale as ${\rm Re}^{0.59}$ under the continuous boundary conditions, while it scales as ${\rm Re}^{0.19}$ under the constant Dirichlet boundary condition.
The present numerical method remains valid when considering solute absorption by a capsule or heating/cooling of a sphere.
\section{Acknowledgments}
The authors thank the Ministère de l’Enseignement Supérieur, de la Recherche et de l’Innovation (MESRI), the Biomechanics and Bioengineering Laboratory (BMBI) and the Agence Nationale de la Recherche (ANR-20-CE45-0008-01) for financial support.
\section*{Appendix: Numerical method}
\label{sec:numerical_method}
\subsection{Lattice Boltzmann method}
Both the flow and the solute advection-diffusion are computed using the lattice Boltzmann method (LBM) \cite{Wolf-Gladrow2000,Succi2001,Sukop2006,Mohamad2011,Kruger2016}, which is based on the lattice Boltzmann equation \eqref{eq:LBE} that gives the evolution in time and space of a distribution function $\chi_i$,
\begin{equation}
\chi_i({\bf r} + {\bf e}_i, t + 1) - \chi_i({\bf r},t) = \Omega_i,
\label{eq:LBE}
\end{equation}
where $\bf{r}$ is the discrete position, ${\bf e}_i$ is the $i$th discrete velocity direction, $t$ is the time, and $\Omega_i$ the collision operator.
In this study, $\Omega_i$ is approximated by the Bhatnagar-Gross-Krook (BGK) collision operator \cite{Bhatnagar1954},
\begin{equation}
\Omega_i = -\frac{\chi_i - \chi_i^{\rm{eq}}}{\tau},
\end{equation}
with $\chi_i^{\rm{eq}}$ the equilibrium distribution function, and $\tau$ the microscopic characteristic relaxation time.

The present studied problem is multiphysics.
It couples fluid flow, mass transfer and fluid-structure interaction.
It requires solving two lattice Boltzmann equations: one for the flow, and the other for the concentration field.
This two-component LBM approach has already been successfully used by the authors in Refs.~\cite{Kaoui2017,Kaoui2018,Kaoui2020,Bielinski2021,Ferrari2012}.
The associated distribution functions for the flow and mass transfer parts are respectively $f_i$ and $g_i$.
Here, the D3Q19 lattice is opted with $i = 0 - 18$.

\begin{figure*}
\centering
\subfloat[\label{subfig:meshes}]{\includegraphics[height = 0.2\textwidth]{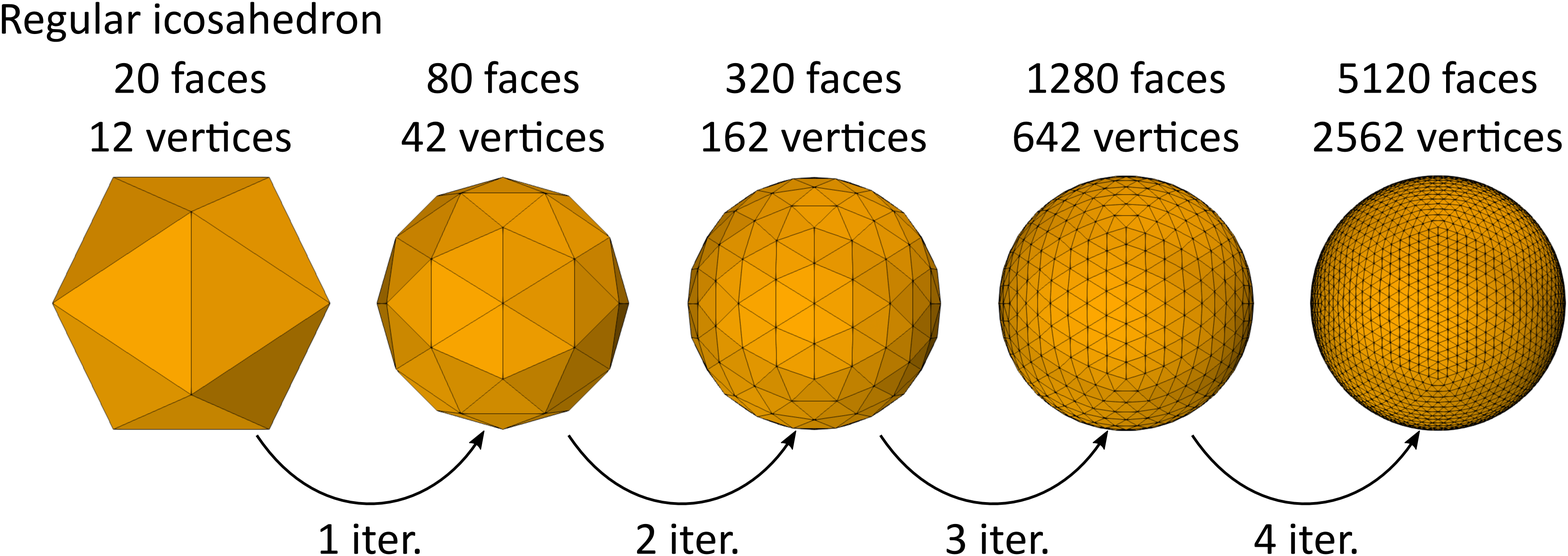}}
\hspace{1.5cm}
\subfloat[\label{subfig:spring_network}]{\includegraphics[height = 0.2\textwidth]{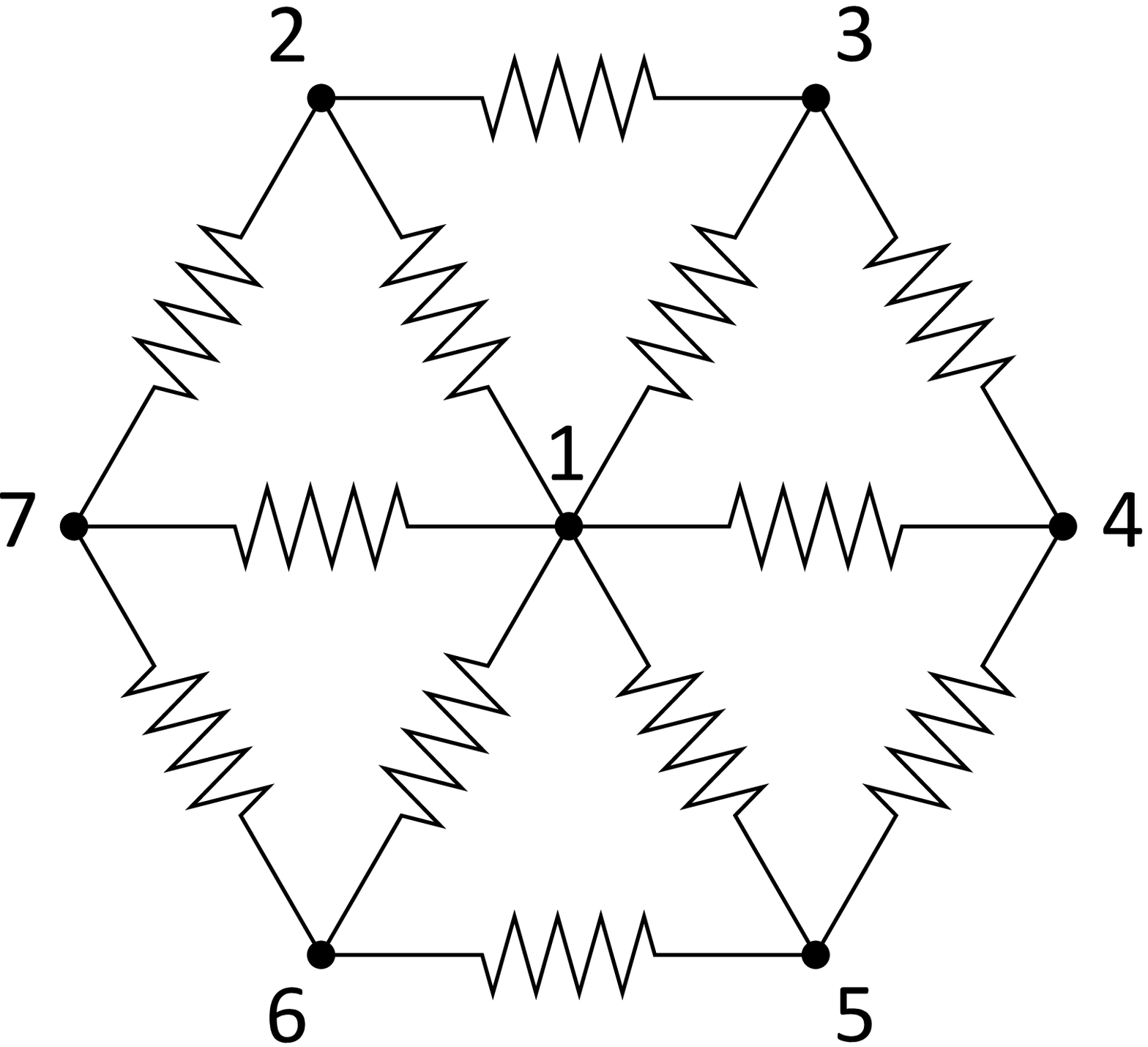}}
\caption{(a) Meshes of the capsule obtained after successive subdivisions of a regular icosahedron, corresponding to icospheres with 80, 320, 1280 and 5120 triangular faces. (b) Spring network connecting the Lagrangian nodes of 6 adjacent faces.}
\label{fig:meshes_springs}
\end{figure*}
The distribution function $f_i$ represents physically the probability to find a population of pseudo-fluid particles at a position $\bf{r}$ at time $t$, flowing at the discrete velocity ${\bf e}_i$.
The lattice Boltzmann equation associated to $f_i$ is,
\begin{equation}
f_i({\bf r} + {\bf e}_i, t + 1) - f_i({\bf r},t) = -\frac{f_i - f_i^{\rm eq}}{\tau_\nu} + \frac{\omega_i}{c_{\rm s}^2}\left({\bf f} \cdot {\bf e}_i\right),
\label{eq:LBE_fi}
\end{equation}
where ${\bf f}$ is the body force exerted on the fluid.
For the D3Q19 lattice, the lattice speed of sound is $c_{\rm s} = 1/\sqrt{3}$, and the weight factors $\omega_i$ are 1/3 for $i = 0$, 1/18 for $i = 1 - 6$ and 1/36 for $i = 7-18$.
The relaxation time $\tau_\nu$ is related to the fluid kinematic viscosity via the relationship,
\begin{equation}
\nu = c_{\rm s}^2\left(\tau_\nu - \frac{1}{2}\right),
\end{equation}
and the equilibrium distribution function $f_i^{{\rm eq}}$ is given by,
\begin{equation}
f_i^{\rm eq} = \omega_i \rho\left(1 + \frac{{\bf e}_i\cdot\bf{u}}{c_{\rm s}^2} + \frac{\left({\bf e}_i \cdot \bf{u}\right)^2}{2c_{\rm s}^4} - \frac{\bf{u} \cdot \bf{u}}{2c_{\rm s}^2}\right).
\end{equation}
The local mass density $\rho$ and the local velocity $\mathbf{u}$ are respectively computed as the zeroth and the first order moments of $f_i$,
\begin{align}
&\rho({\bf r},t) = \sum _{i=0}^{18} f_i ({\bf r},t),\\
&{\bf u}({\bf r},t) = \frac{1}{\rho({\bf r},t)}\sum _{i=0}^{18} f_i ({\bf r},t){\bf e}_i,
\end{align}
and the local hydrodynamic pressure is given by the equation of state $p = \rho c_{\rm s}^2$.
The upper and the lower walls are translated in opposite directions to generate shear flow using the Ladd's bounce-back boundary condition \cite{Ladd1994},
\begin{equation}
f_{-i}({\bf r},t + 1) = f_i^*({\bf r},t) - 6\omega_i\rho\frac{{\bf e}_i\cdot {\bf u}_{\rm w}}{c_{\rm s}},
\end{equation}
where ${\bf e}_{-i} = - {\bf e}_i$, ${\bf u}_{\rm w}$ is the desired wall velocity such as ${\bf u}_{\rm w} = (U,0,0)$ at the top wall and ${\bf u}_{\rm w} = (-U,0,0)$ at the bottom wall.
The superscript $*$ refers to the postcollision state.
For the mass transfer part, the corresponding lattice Boltzmann equation is,
\begin{equation}
g_i({\bf r} + {\bf e}_i, t + 1) - g_i({\bf r},t) = -\frac{g_i - g_i^{\rm{eq}}}{\tau_D},
\label{eq:LBE_gi}
\end{equation}
where the relaxation time $\tau_D$ is related to the solute diffusion coefficient by,
\begin{equation}
D = c_{\rm s}^2\left(\tau_D - \frac{1}{2}\right),
\end{equation}
and the corresponding equilibrium distribution function is,
\begin{equation}
g_i^{\rm eq} = \omega_i c\left(1 + \frac{{\bf e}_i\cdot\bf{u}}{c_{\rm s}^2} + \frac{\left({\bf e}_i \cdot \bf{u}\right)^2}{2c_{\rm s}^4} - \frac{\bf{u} \cdot \bf{u}}{2c_{\rm s}^2}\right).
\end{equation}
The local concentration is computed as the zeroth order moment of the distribution function $g_i$,
\begin{equation}
c({\bf r},t) = \sum _{i=0}^{18} g_i ({\bf r},t).
\end{equation}
Zero concentration is set at the simulation box edges using the Zhang's bounce-back boundary condition \cite{Zhang2012},
\begin{equation}
g_{-i}({\bf r},t + 1) = -g_i^*({\bf r},t) + 2\omega_ic_{\rm w},
\label{eq:Zhang}
\end{equation}
where $c_{\rm w}$ is the desired wall concentration.
This scheme has also been used to run simulations with constant concentration at the capsule surface, as largely used in the literature and for comparison purpose with the unsteady continuous boundary conditions studied in the present article.
The BGK relaxation times in the lattice Boltzmann equations \eqref{eq:LBE_fi} and \eqref{eq:LBE_gi} are set to $\tau_{\nu} = 1$ and $\tau_{D} = 0.55$, respectively.
These values are carefully chosen to ensure both the accuracy of the computed solution and the stability of the numerical scheme.
\subsection{Immersed boundary method}
The surrounding flowing fluid exerts a force on the membrane of the capsule, which, in turn, exerts a reactive force back on the fluid.
The two-way coupling between the fluid flow and the capsule dynamics is achieved with the immersed boundary method (IBM) \cite{Peskin1977}.
The capsule membrane is approximated by a set of marker points $\{ {\bf r}_j(t)\}$ that are distributed on a Lagrangian mesh.
The Lagrangian mesh is generated by performing successive subdivisions of a regular icosahedron \cite{script_icosphere}, as illustrated in Fig.~\ref{subfig:meshes}.
At each iteration, the edges of the polyhedron are split in their midpoint, and the newly created vertices are radially projected on the circumscribed sphere of desired radius $R$.
This process ensures the symmetry of the mesh.
The mechanics of the capsule's membrane is accounted for by a spring network connecting the Lagrangian markers, as represented in Fig.~\ref{subfig:spring_network}.
The velocity of the Lagrangian markers $\dot{{\bf r}}_j(t)$ is interpolated from the velocity field ${\bf u}({\bf r},t)$ computed with the LBM on the Eulerian mesh grid,
\begin{equation}
\dot{{\bf r}}_j(t)  = \sum_{\bf r} {\bf u}({\bf r},t)\Delta({\bf r}_j(t),{\bf r}),
\end{equation}
where $\Delta$ is a smooth approximation of the Dirac distribution function given by $\Delta({\bf r}_1,{\bf r}_2) = \delta(x_1,x_2)\delta(y_1,y_2)\delta(z_1,z_2)$ with,
\begin{equation}
\delta \left(a,b\right) = \begin{cases} \frac{1}{4}\left(1+ \cos \frac{\pi \left(a - b\right)}{2}\right)~~~\text{if}~\left|a - b \right| \leq 2 \\
0 ~~~\text{else}
\end{cases}
\text{.}
\end{equation}
The new positions of the Lagrangian markers at time $t+1$ are obtained using the explicit Euler scheme,
\begin{equation}
{\bf r}_j(t+1) = {\bf r}_j(t) + \dot{{\bf r}}_j(t).
\end{equation}
After the advection step, the Lagrangian markers exert a force back on the fluid.
The reactive force exerted on the Eulerian mesh is obtained by extrapolation from the Lagrangian mesh using again the function $\Delta$,
\begin{equation}
{\bf f}({\bf r},t) = \sum_j {\bf f}_j(t)\Delta({\bf r}_j(t),{\bf r})\text{,}
\label{eq:force}
\end{equation}
${\bf f}_j$ being the force acting at the Lagrangian marker $j$.
It is computed as the sum of the spring forces acting on the node $j$, see Fig.~\ref{subfig:spring_network},
\begin{equation}
{\bf f}_j = -\sum_{i\neq j} \kappa \left( \lVert{\bf r}_{ij}\rVert - \lVert{\bf r}^0_{ij}\rVert\right) \frac{{\bf r}_{ij}}{\lVert{\bf r}_{ij}\rVert},
\label{eq:spring_force}
\end{equation}
with $\kappa$ the spring constant and ${\bf r}_{ij} = {\bf r}_i - {\bf r}_j$.
The summation is over nodes $i$ that are connected to the node $j$ by a spring.
Superscript ``$0$'' refers to the initial undeformed state.
The force density ${\bf f}$ is then included into the lattice Boltzmann equation \eqref{eq:LBE_fi} to compute the new flow field, modified by the capsule dynamics.
The capsule mesh is an icosphere with 320 triangular faces and 162 vertices (see Fig.~\ref{subfig:meshes}).
The spring constant is set to $\kappa = 7$ to model a nondeformable spherical capsule.
\subsection{Validation of the numerical method}
\label{sec:validation}
\begin{figure}
\centering
\includegraphics[width = 0.4\textwidth]{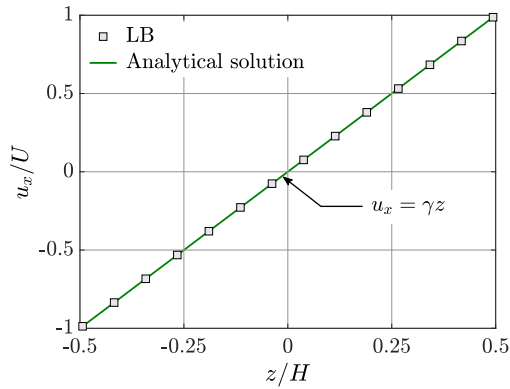}
\caption{Fluid velocity profile computed by the LBM (squares) along with the analytical solution of the linear shear flow (solid line). Excellent agreement is achieved between the LBM result and the analytical solution, which validates the flow solver.}
\label{fig:comp_LB_analytical_shear_flow}
\end{figure}

The validation of the numerical method consists of three parts: the flow solver, the fluid-structure interaction solver, and the mass transfer solver.
First, the flow solver is validated by comparing the velocity profile computed in absence of the capsule to the expected analytical solution for shear flow in the $x$-direction,
\begin{equation}
{\bf u} = \left(\gamma z,0,0\right).
\end{equation}
Figure \ref{fig:comp_LB_analytical_shear_flow} shows perfect match between the obtained numerical solution and theory that confirms that the implementation of the LBM flow solver is correct.

The next step in the validation process concerns the fluid-structure coupling part, which is achieved by the IBM.
For this, a capsule is suspended in a fluid subjected to simple shear.
The resulting angular velocity $\omega$ of the capsule is measured for various applied shear rates $\gamma$.
The latter is varied by varying solely the wall velocity $U$ from $0.01$ to $0.05$, while holding all the other parameters constant.
For an unbounded domain, and in the Stokes flow limit ($\mathrm{Re} \ll 1$), the angular velocity $\omega$ of a rigid sphere rotating due to simple shear flow is \cite{Clift1978}, $\omega = \gamma / 2$.
Hereafter, it is explained how this quantity is computed in the present work.
\begin{figure}[t]
\centering
\subfloat[\label{subfig:snapshots_capsule_rotation}]
{\includegraphics[width=0.45\textwidth]{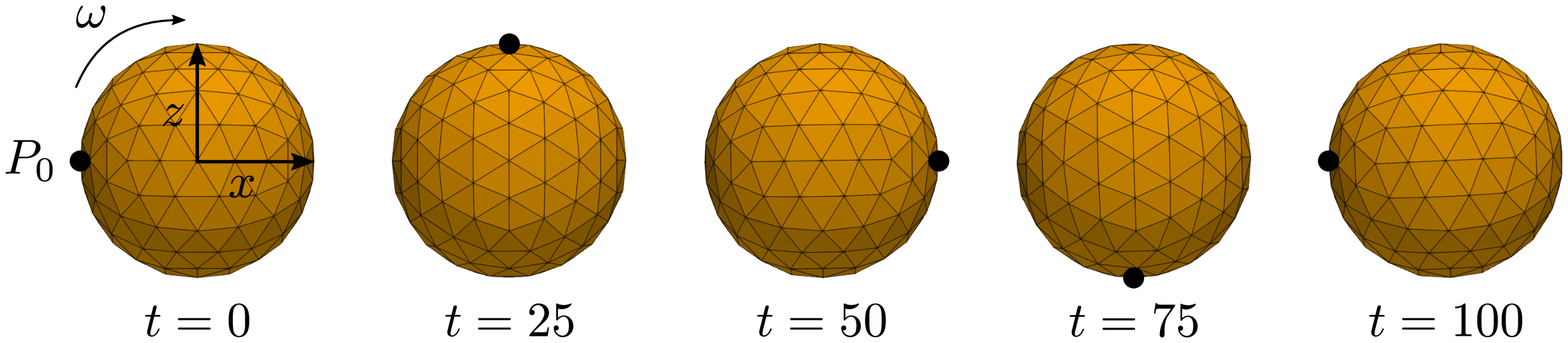}}\\
\subfloat[\label{subfig:normalized_position}]
{\includegraphics[width=0.5\textwidth]{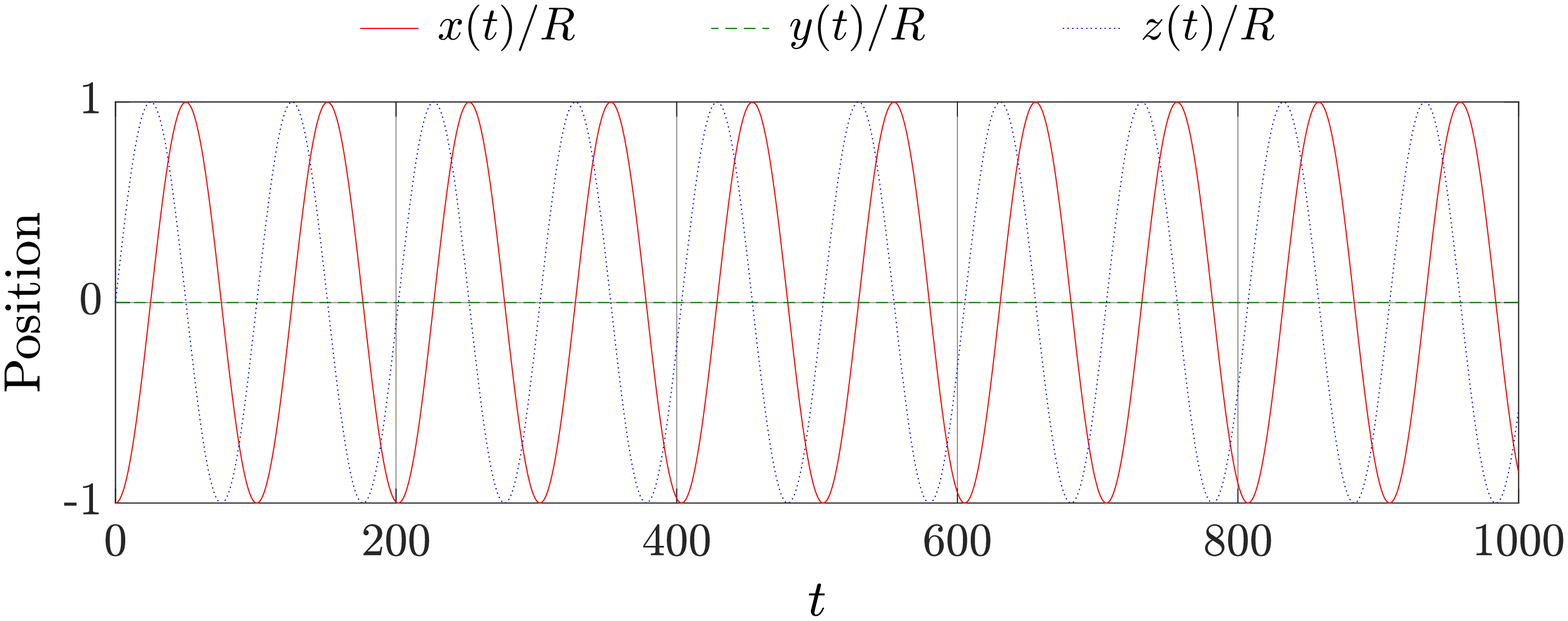}}\\
\subfloat[\label{subfig:theta_vs_t}]
{\includegraphics[width=0.5\textwidth]{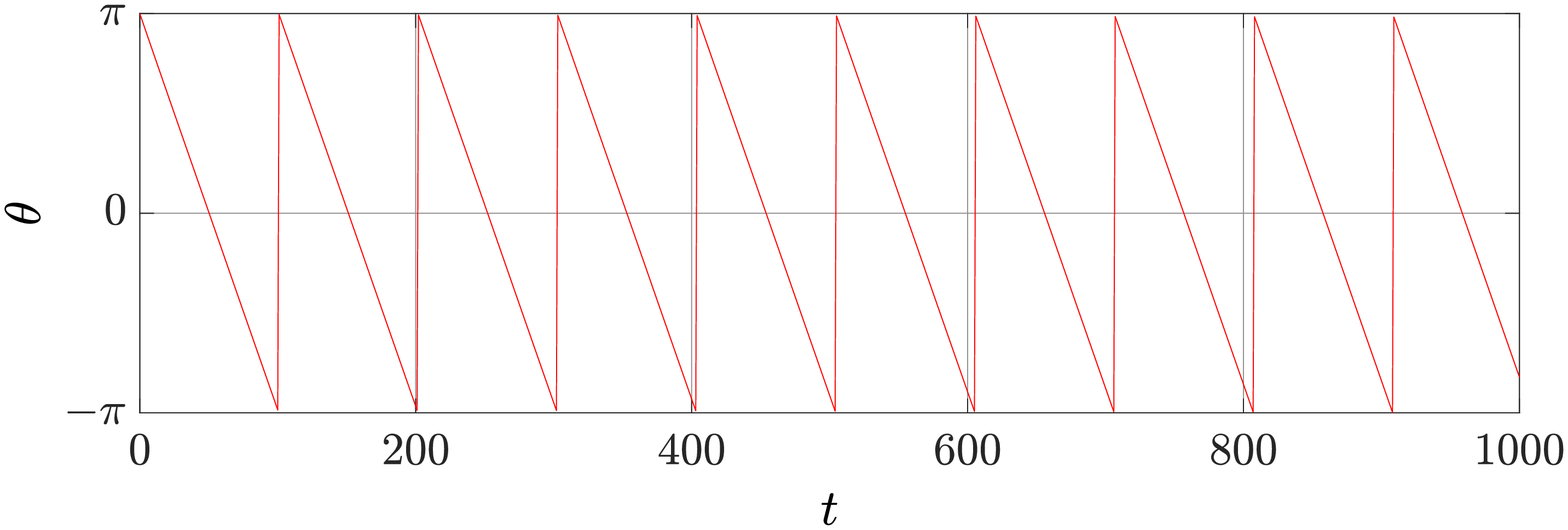}}
\caption{(a) Snapshots showing the rotation of a spherical capsule under shear flow. (b) Normalized position of the point $P_0$ initially located at ${\bf r} = (-R,0,0)$ on the capsule membrane. (c) The instantaneous angular position $\theta$ of this point.
The capsule rotates around the $y$-axis at an angular velocity $\omega = \dot \theta = 6.227\times10^{-4}$ l.u. for the applied wall velocity of $U = 0.05$ l.u. (LBM lattice units).
The figures demonstrate the robustness and the numerical stability of the computer code when the capsule undergoes multiple rotational cycles.
}
\label{fig:IBM_validation}
\end{figure}
\begin{table}[b]
\centering
\begin{tabular}{c c c c c c}
\hline
$U$ & $\mathrm{Re}$ (10$^{-1}$) & $\gamma$ (10$^{-4}$) & $\omega$ (10$^{-4}$) & $\omega/\gamma$ & ${\rm E}$ (\%)\\ [0.5ex]
\hline
%\hline
0.01 & 0.97 & 2.532 & 1.259 & 0.497 & 0.6\\
%\hline
0.02 & 1.94 & 5.063 & 2.515 & 0.497 & 0.6\\
%\hline
0.03 & 2.92 & 7.595 & 3.762 & 0.495 & 1.0\\
%\hline
0.04 & 3.89 & 10.13 & 5.000 & 0.494 & 1.2\\
%\hline
0.05 & 4.86 & 12.66 & 6.227 & 0.492 & 1.6\\
\hline
\end{tabular}
\caption{Angular velocity $\omega$ of a capsule under shear flow at various wall velocities $U$.
The ratios $\omega/\gamma$ are very close to the expected theoretical value $1/2$, which validates the implementation of the fluid-structure coupling. The relative error ${\rm E}$ does not exceed $2\%$ for the highest Reynolds number.
$U$, $\gamma$ and $\omega$ are expressed in LBM lattice units.}
\label{tab:omega_gamma}
\end{table}
Figure \ref{subfig:normalized_position} represents the dynamics of a point $P_0$ on the membrane of the capsule that is initially located at ${\bf r} = (-R,0,0)$.
For these simulations the spring stiffness is set to $\kappa = 7$, which is large enough so that the capsule does not deform.
Thus, it remains spherical.
This allows to recover the analytical solution, which is derived originally for a spherical rigid particle.
The $x$ and $z$ coordinates of the point undergo regular oscillations, while the $y$ coordinate stays at 0, indicating the capsule does not translate along the $y$-axis.
This confirms that the mesh remains symmetric and the computer code is robust and numerically stable after many cycles of rotations.
Similar results, but with different oscillations frenquencies are obtained at all other wall velocities.
The instantaneous angular position of the point is then computed as $\theta = \arctan\left(\frac{z}{x}\right)$, and it is reported in Fig.~\ref{subfig:theta_vs_t}.
This figure shows periodic oscillations in time meaning the capsule rotates at constant angular velocity $\omega = \dot \theta$.
The obtained angular velocities $\omega$ at various flow shear rates $\gamma$, and the ratios $\omega/\gamma$ are reported in Tab.~\ref{tab:omega_gamma}.
The computed ratios $\omega/\gamma$ are all very close to the expected theoretical value $1/2$, which means the fluid-structure interaction part is well resolved.
The slight deviation from $1/2$ (that never exceeds $2\%$) observed when increasing the Reynolds number is explained by the fact that the theory is valid only in the Stokes flow limit (${\rm Re \ll 1}$) and for an unbounded domain.

\begin{figure}
\centering
\includegraphics[width = 0.4\textwidth]{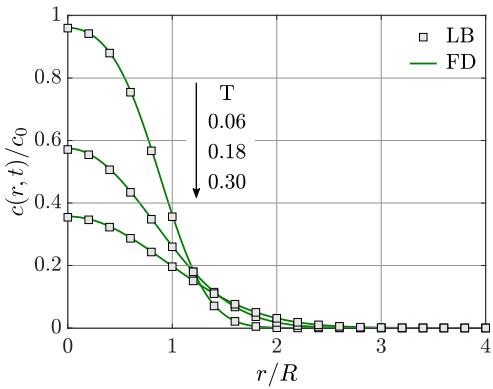}
\caption{Computed concentration profiles by LBM (squares) and FDM (solid lines) in the case of solute release from a spherical capsule in a fluid at rest.}
\label{fig:comp_LB_FD_concentration_profile}
\end{figure}

Finally, the mass transfer part is validated by comparing the concentration profile obtained by LBM with the solution computed by the finite difference method (FDM), in the case of solute release from a capsule in the absence of flow (${\rm Re = 0}$).
For this case, the diffusion is purely radial and the solution adopts a spherical symmetry.
Figure \ref{fig:comp_LB_FD_concentration_profile} shows both the LBM and FDM normalized concentration profiles computed at various dimensionless times.
Both solutions are in excellent agreement, and thus, validating the mass transfer solver.
\end{document}